\documentclass[journal]{IEEEtran}
\usepackage{cite}
\usepackage[pdftex]{graphicx}
\usepackage{amsmath}
\usepackage{amssymb}
\usepackage{algorithm}
\usepackage[noend]{algpseudocode}
\usepackage{mathtools}
\usepackage{hyperref}
\usepackage{color}

\newtheorem{ass}{Assumption}
\newtheorem{thm}{Theorem}
\newtheorem{crl}{Corollary}
\newtheorem{prop}{Proposition}
\newtheorem{lem}{Lemma}
\newtheorem{rem}{Remark}

\renewcommand{\baselinestretch}{0.93}

\begin{document}

\title{Linear Reduced Order Model Predictive Control}

\author{Joseph~Lorenzetti, Andrew~McClellan, Charbel~Farhat, Marco~Pavone% <-this % stops a space
\thanks{The authors are with the Department of Aeronautics and Astronautics, Stanford University, Stanford, CA, 94305 USA: (\{jlorenze, amcclell, cfarhat, pavone\}@stanford.edu). This work was supported by the Office of Naval Research  (Grant N00014-17-1-2749).}}

% The paper headers
%\markboth{Journal of \LaTeX\ Class Files,~Vol.~14, No.~8, August~2015}%
%{Shell \MakeLowercase{\textit{et al.}}: Bare Demo of IEEEtran.cls for IEEE Journals}
% The only time the second header will appear is for the odd numbered pages
% after the title page when using the twoside option.
% 
% *** Note that you probably will NOT want to include the author's ***
% *** name in the headers of peer review papers.                   ***
% You can use \ifCLASSOPTIONpeerreview for conditional compilation here if
% you desire.

% make the title area
\maketitle

% As a general rule, do not put math, special symbols or citations
% in the abstract or keywords.
\begin{abstract}
Model predictive controllers use dynamics models to solve \textit{constrained} optimal control problems. However, computational requirements for real-time control have limited their use to systems with low-dimensional models. 
Nevertheless, high-dimensional models arise in many settings, for example discretization methods for generating finite-dimensional approximations to partial differential equations can result in models with thousands to millions of dimensions.
In such cases, reduced order models (ROMs) can significantly reduce computational requirements, but model approximation error must be considered to guarantee controller performance.
In this work, a reduced order model predictive control (ROMPC) scheme is proposed to solve robust, output feedback, constrained optimal control problems for high-dimensional linear systems. Computational efficiency is obtained by using projection-based ROMs, and guarantees on robust constraint satisfaction and stability are provided. Performance of the approach is demonstrated in simulation for several examples, including an aircraft control problem leveraging an inviscid computational fluid dynamics model with dimension 998,930.
\end{abstract}

% Note that keywords are not normally used for peerreview papers.

% For peer review papers, you can put extra information on the cover
% page as needed:
% \ifCLASSOPTIONpeerreview
% \begin{center} \bfseries EDICS Category: 3-BBND \end{center}
% \fi
%
% For peerreview papers, this IEEEtran command inserts a page break and
% creates the second title. It will be ignored for other modes.
\IEEEpeerreviewmaketitle

\section{Introduction}

\IEEEPARstart{M}{odel} predictive control (MPC) \cite{RawlingsMayneEtAl2017} is a powerful technique for addressing \textit{constrained} optimal control problems, and has been widely studied in both theory and practice. MPC algorithms compute optimal control sequences in a receding horizon fashion by solving (typically online) constrained mathematical programs that optimize the system's predicted future trajectory over a finite horizon. The predictive capability is obtained by embedding a mathematical model of the system's dynamics as constraints in the optimization, and additional state and control constraints can also be included.

While MPC algorithms offer a number of advanced capabilities beyond classical control techniques, they are also burdened by unique challenges. One significant and fundamental challenge is the trade-off between the computational complexity of the optimization and the performance of the closed-loop system. In particular, the receding horizon nature of MPC requires solving the optimization problem in real-time, which precludes the use of overly complex dynamics models. Contrarily, high-fidelity dynamics models are required for good prediction accuracy and thus good closed-loop performance.
Many early applications of MPC were for industrial process control where the system dynamics are slow, allowing for a slow controller update rate and thus the accommodation of higher fidelity models. More recently, increased computational capabilities and more advanced optimization algorithms have led to the application of MPC to systems with faster dynamics, such as automotive \cite{BealGerdes2013} or aerospace \cite{ErenPrachEtAl2017} vehicles. However in these cases the dynamics models are generally low-dimensional.

This work explores a computationally efficient MPC approach for linear systems whose models are high-dimensional. 
This problem setting is particularly motivated by cases where the high-dimensional model is defined by a discretization (e.g. by a finite element or finite volume method) of a partial differential equation (PDE) model of an infinite-dimensional system.
Such models arise in many engineering problems for modeling structural or fluid dynamics, heat flow dynamics, chemical reactions, electrochemical processes, and more. Example control applications involving these types of systems include soft robot control \cite{ThieffryKruszewskiEtAl2019, TonkensLorenzettiEtAl2021}, aircraft control \cite{LorenzettiMcClellanEtAl2020}, fluid flow control \cite{Lassaux2002}, chemical reaction processes \cite{AgudeloEspinosaEtAl2007}, lithium-ion battery charging \cite{FanLiEtAl2018}, among many others. In these settings, it is not uncommon for the model dimension of the discretized PDE to reach from the thousands to millions, making standard model-based controller design (e.g. linear quadratic regulation, MPC) extremely challenging \cite{Benner2004}.
In this context, one solution to the problem of computational complexity is to design controllers based on \textit{low-dimensional} surrogate models. However, the use of low-fidelity models for controller design can lead to poor closed-loop performance, including sub-optimality, poor robustness to disturbances, or even instability.  Therefore a \textit{low-dimensional} but \textit{high-fidelity} model is required.

In fact, principled model order reduction techniques for deriving high-fidelity reduced order models (ROMs) from high-dimensional models have been extensively developed \cite{Antoulas2005, BennerGugercinEtAl2015}. These model reduction methods, such as balanced truncation and proper orthogonal decomposition (POD), have been successfully applied for model-based control of infinite-dimensional systems. For example, POD is used in \cite{MarquezEspinosaOviedoEtAl2013} for controlling a nuclear reactor and in \cite{McClellanLorenzettiEtAl2020} for aircraft control. Performance of control systems based on reduced order models has also been analyzed from a theoretical perspective, including for the unconstrained linear quadratic optimal control problem \cite{GubischVolkwein2017, AntilHeinkenschlossEtAl2010, Zhang2019}, as well as in receding horizon control \cite{GhiglieriUlbrich2014, AllaVolkwein2015, Altmueller2014}. While these works provide some theoretical results, such as bounds on the sub-optimality of the controller or guaranteed stability of the high-dimensional model, they do not consider optimal control problems with \textit{state constraints}, which are of considerable practical interest. The unique challenge of using reduced order models for state-constrained MPC is that constraint violations by the controlled high-dimensional system may occur due to model approximation error.

Previous work has begun to address the problem of state-constrained MPC with reduced order models. Early work \cite{HovlandGravdahlEtAl2008, HovlandLovaasEtAl2008} only considers soft state constraints, and therefore no guarantees on constraint satisfaction are provided. Other approaches provide hard constraint satisfaction guarantees by appropriately analyzing the model reduction error. In some work, certain conditions are required to provide stability and constraint satisfaction guarantees, for example by restricting the analysis to ROMs derived using modal decomposition \cite{DubljevicEl-FarraEtAl2006} or to problems where the constraints are only imposed on the reduced order state \cite{SopasakisBernardiniEtAl2013}. A less restrictive approach that considers projection-based ROMs and general state and control constraints is described in \cite{LoehningRebleEtAl2014}. However this work uses full-state feedback to ensure the model reduction error dynamics are stable, and for high-dimensional systems it is often not practical to assume that full-state measurements are available. This limitation is addressed in \cite{KoegelFindeisen2015b}, which provides stability and constraint satisfaction guarantees while incorporating a reduced order state estimator into an output-feedback control scheme. Additionally, this work considers a robust control problem with bounded disturbances on the dynamics and measurement noise. The disadvantage of this approach is that the error analysis requires the solution to linear programs whose complexity is dependent on the dimension of the full order model. Thus for extremely high-dimensional problems this approach becomes less computationally efficient.

In summary, traditional MPC methods rely on computational tools that do not scale to extremely high-dimensional problems. While previous work has introduced reduced order models within the context of MPC to overcome online computational challenges, no work has \textit{simultaneously} provided a method that (1) has theoretical performance guarantees, (2) is computationally efficient to synthesize, (3) is applicable to output feedback settings, and (4) can leverage general projection-based model order reduction methods; all are required or desirable properties for practical control settings.

{\em Statement of Contributions:} In this work we present a reduced order model predictive control (ROMPC) scheme for controlling high-dimensional discrete-time linear systems. In particular, we consider a robust, output feedback, constrained optimal control problem where bounded disturbances affect the system dynamics and measured outputs, and where constraints are imposed on both states and control inputs. The proposed ROMPC scheme is defined by a constrained optimization problem that leverages a reduced order model, an ancillary feedback controller, and a reduced order state estimator. An efficient method for synthesizing the ancillary controller and estimator gains (based on $\mathcal{H}_2$-optimal control techniques) is presented, and an efficient approach for analyzing errors associated with the model order reduction, state estimation, and exogenous disturbances based on linear programming is presented.
Theoretical guarantees on the stability of the closed-loop system and robust constraint satisfaction are also provided. Furthermore, discussion on the application of the proposed ROMPC scheme to setpoint tracking problems and also to high-dimensional continuous-time systems is presented. 
This work\footnote{ {https://github.com/StanfordASL/rompc}} is based on our preliminary works \cite{LorenzettiLandryEtAl2019, LorenzettiPavone2020b} and contains additional contributions related to the design of the ancillary controller and estimator gains, as well as higher-dimensional examples including an aircraft control problem leveraging a computational fluid dynamics (CFD) model with one million degrees of freedom.

{\em Organization:}
We begin in Section \ref{sec:prob} by defining the control problem and the full and reduced order models. In Section \ref{sec:rompc} we define the ROMPC scheme, and then prove results on robust constraint satisfaction and stability in Section \ref{sec:closedloopperf}. An efficient methodology for synthesizing the controller is proposed in Section \ref{sec:controllersynth}, and an efficient methodology for error analysis that is used to guarantee robust constraint satisfaction is included in Section \ref{sec:errorbounds}. We then provide a discussion on the application of ROMPC for setpoint tracking in Section \ref{sec:setpointtracking} and for continuous-time problems in Section \ref{sec:continuoustime}. Section \ref{sec:experiments} is dedicated to demonstrating the performance of the proposed method through simulation.

\section{Problem Formulation} \label{sec:prob}
This section defines the system dynamics model, the mathematical formulation for the control problem, and the nominal reduced order model used to design the ROMPC scheme.

\subsection{Full Order Model}
This work considers high-dimensional system models, such as from a finite approximation to an infinite-dimensional system (e.g. a discretized PDE). Specifically, consider the linear, discrete-time, full order model (FOM)
\begin{equation} \label{eq:fom}
\begin{split}
x^f_{k+1} &= A^fx^f_k + B^fu_k + B^f_w w_k, \\
y_k &= C^f x^f_k + v_k, \quad z_k = H^f x^f_k,\\
\end{split}
\end{equation}
where $(\cdot)^f$ denotes a full order variable, $x^f \in \mathbb{R}^{n^f}$ is the state, $u \in \mathbb{R}^m$ is the control, $y \in \mathbb{R}^p$ is the measurement, $z \in \mathbb{R}^o$ are performance variables, and $A^f \in \mathbb{R}^{n^f \times n^f}$, $B^f \in \mathbb{R}^{n^f \times m}$, $B^f_w \in \mathbb{R}^{n^f \times m_w}$, $C^f \in \mathbb{R}^{p \times n^f}$, and $H^f \in \mathbb{R}^{o \times n^f}$ are the system matrices. The disturbances $w \in \mathbb{R}^{m_w}$ and $v \in \mathbb{R}^{p}$ are assumed to be bounded by
\begin{equation} \label{eq:noise}
w \in \mathcal{W}, \quad v \in \mathcal{V},
\end{equation}
where $\mathcal{W} \coloneqq \{w \: | \: H_w w \leq b_w \}$ with $H_w \in \mathbb{R}^{n_w \times m_w}$ and $\mathcal{V} \coloneqq \{v \: | \: H_v v \leq b_v \}$ with $H_v \in \mathbb{R}^{n_v \times p}$ are convex polytopes.

Performance and control constraints are defined by
\begin{equation} \label{eq:constraints}
z \in \mathcal{Z}, \quad u \in \mathcal{U}, 
\end{equation}
where $\mathcal{Z} \coloneqq \{z\:|\:H_z z \leq b_z \}$ with $H_z \in \mathbb{R}^{n_z \times o}$ and $\mathcal{U} \coloneqq \{u\:|\:H_u u \leq b_u \}$ with $H_u \in \mathbb{R}^{n_u \times m}$ are also convex polytopes. The following assumption is also made regarding the constraints and disturbances:

\begin{ass} \label{ass:compact}
The sets $\mathcal{Z}$, $\mathcal{U}$, $\mathcal{W}$, and $\mathcal{V}$ are compact and $\mathcal{Z}$ and $\mathcal{U}$ contain the origin in their interior.
\end{ass}

The performance variables $z$ are defined because in many practical high-dimensional problems only a small subset of the states are relevant from a performance perspective. For example in rigid-body aircraft control problems leveraging CFD aerodynamics models (i.e. a fluid-structure interaction problem), the variables of interest are the aircraft rigid-body states (e.g. position, velocity, attitude) and not the fluid states.

\subsection{Constrained Optimal Control Problem}
The control problem is to find an output feedback control scheme that guarantees that the FOM \eqref{eq:fom} satisfies the performance and control constraints \eqref{eq:constraints} under any admissible disturbances, is stable, and also minimizes the cost function
\begin{equation} \label{eq:fomcost}
J = \sum_{k=0}^\infty (x^f_k)^T Q^f x^f_k + u_k^T R u_k,
\end{equation}
where $Q^f \in \mathbb{R}^{n^f \times n^f}$ is symmetric, positive semi-definite and $R \in \mathbb{R}^{m \times m}$ is symmetric, positive definite.

\subsection{Reduced Order Model}
Typical robust output feedback MPC schemes \cite{RawlingsMayneEtAl2017, LorenzettiPavone2020} would leverage the dynamics model \eqref{eq:fom} to solve the constrained optimal control problem. But, with a high-dimensional model the resulting computational requirements may be excessive. An alternative is to use a reduced order model to design a computationally efficient controller. In fact, model order reduction methods \cite{Antoulas2005, BennerGugercinEtAl2015} have been successfully used to derive low-order (but high-fidelity) models for many interesting high-dimensional systems, and in particular for those arising from finite approximations of infinite-dimensional systems. 

In this work we consider ROMs derived using projection-based model reduction methods, a general class of methods that includes popular techniques such as balanced truncation and proper orthogonal decomposition. These methods utilize either a Galerkin or a Petrov-Galerkin projection to project the model onto a reduced order subspace. In particular, a pair of basis matrices $V,W \in \mathbb{R}^{n^f \times n}$ define the projection matrix\footnote{If $W=V$ the projection is referred to as a Galerkin projection, and some model reduction methods define the basis such that $W^TV = I$.} $P = V(W^TV)^{-1}W^T$, and the high-dimensional state can be projected to the reduced order state by $x = (W^TV)^{-1}W^Tx^f$ and can be approximately reconstructed by $x^f \approx Vx$. 

The reduced order model is therefore defined using the Petrov-Galerkin projection by
\begin{equation} \label{eq:orig_rom}
\begin{split}
x_{k+1} &= Ax_k + Bu_k + B_w w_k, \\
y_k &= C x_k + v_k, \quad z_k = H x_k, \\
\end{split}
\end{equation}
where $x \in \mathbb{R}^{n}$ is the reduced order state and the ROM dynamics matrices are defined by $A \coloneqq (W^TV)^{-1}W^TA^fV$, $B \coloneqq (W^TV)^{-1}W^TB^f$, $B_w \coloneqq (W^TV)^{-1}W^TB^f_w$, $C \coloneqq C^fV$, and $H \coloneqq H^fV$. We also make the following assumption:
\begin{ass} \label{ass:ctrlobsv}
The pair $(A, B)$ is controllable and the pairs $(A,C)$ and $(A,H)$ are observable.
\end{ass}
As will be seen later, this assumption is required to guarantee performance of the ROMPC algorithm, and will also be used to bound errors induced by the model approximation in Section \ref{sec:errorbounds}. Since this assumption is based on the lower dimensional ROM it is computationally easy to verify, but future work could be done to provide conditions on the full order model or model reduction process that guarantee its satisfaction. This could potentially be done using an approach similar to \cite{AmsallemFarhat2012}, which guarantees stability in projection-based ROMs.

Additionally, the cost function \eqref{eq:fomcost} is approximated as
\begin{equation} \label{eq:romcost}
J \approx \sum_{k=0}^\infty x_k^T Q x_k + u_k^T R u_k, \quad Q = V^TQ^fV.
\end{equation}

\section{Reduced Order Model Predictive Control} \label{sec:rompc}
Model-based controller design with high-dimensional models is computationally challenging for both the controller synthesis stage (offline), as well as for online implementation. In this work, the ROM \eqref{eq:orig_rom} is therefore leveraged to design an efficient reduced order model predictive control (ROMPC) scheme that consists of: (1) a \textit{reduced order} linear feedback control law, (2) a \textit{reduced order} state estimator, and (3) a \textit{reduced order} optimal control problem (OCP).
In particular, the proposed approach uses the reduced order OCP to optimize the trajectory of a \textit{simulated} nominal reduced order system
\begin{equation} \label{eq:rom}
\begin{split}
\bar{x}_{k+1} &= A\bar{x}_k + B\bar{u}_k, \\
\bar{z}_k &= H \bar{x}_k, \\
\end{split}
\end{equation}
where $\bar{x} \in \mathbb{R}^{n}$, $\bar{u} \in \mathbb{R}^m$, and $\bar{z} \in \mathbb{R}^o$ are the state, control input, and performance output of the simulated ROM, respectively.
The feedback control law, in concert with the state estimator, then drives the real system to track this optimized (simulated) trajectory.

\subsection{Reduced Order Controller and State Estimator}
The reduced order linear feedback control law that is used to control the full order system is given by
\begin{equation} \label{eq:controller}
u_k = \bar{u}_k + K(\hat{x}_k - \bar{x}_k),
\end{equation}
where $(\bar{x}_k, \bar{u}_k)$ are the state and control values of the simulated ROM, $\hat{x}_k$ is the reduced order state estimate, and $K \in \mathbb{R}^{m \times n}$ is the controller gain matrix. The reduced order state estimator is defined as
\begin{equation} \label{eq:estimator}
\hat{x}_{k+1} = A\hat{x}_k + Bu_k + L(y_k - C\hat{x}_k),
\end{equation}
where $u_k$ and $y_k$ are the control and (noisy) measurement from the full order system \eqref{eq:fom}, and $L \in \mathbb{R}^{n \times p}$ is the estimator gain matrix. Both gain matrices $K$ and $L$ are computed using Algorithm \ref{alg:controllersynth} in Section \ref{sec:controllersynth}.

\subsection{Reduced Order Optimal Control Problem} \label{subsec:roocp}
The optimal control problem (OCP) that defines the trajectory of the simulated ROM optimizes over the sequences $\bar{\mathbf{x}}_k=\{\bar{x}_{i|k} \}_{i=k}^{k+N}$ and $\bar{\mathbf{u}}_k = \{\bar{u}_{i|k} \}_{i=k}^{k+N-1}$, where $\bar{x}_{i|k}$ and $\bar{u}_{i|k}$ denote the state and control variables at time step $i$ in the OCP solved at time step $k$. The OCP is given by
\begin{equation} \label{eq:rompc}
\begin{split}
(\bar{\mathbf{x}}^*_k, \bar{\mathbf{u}}^*_k) = \underset{\bar{\mathbf{x}}_k, \bar{\mathbf{u}}_k}{\text{argmin.}} \:\:& \lVert \bar{x}_{k+N|k} \rVert^2_P + \sum_{j=k}^{k+N-1}\lVert \bar{x}_{j|k}\rVert^2_Q + \lVert \bar{u}_{j|k}\rVert^2_R, \\
\text{subject to} \:\: & \bar{x}_{i+1|k} = A\bar{x}_{i|k} + B\bar{u}_{i|k},  \\
& H\bar{x}_{i|k} \in \bar{\mathcal{Z}}, \quad \bar{u}_{i|k} \in \bar{\mathcal{U}}, \\
& \bar{x}_{k+N|k} \in \bar{\mathcal{X}}_f, \quad \bar{x}_{k|k} = \bar{x}_k,
\end{split}
\raisetag{3\normalbaselineskip}
\end{equation}
where $i=k,\dots,k+N-1$, the integer $N$ is the planning horizon, $\bar{x}_k$ is the current state of the simulated ROM, and the constraint sets $\bar{\mathcal{Z}}$ and $\bar{\mathcal{U}}$ are tightened versions of the original constraints \eqref{eq:constraints} such that $\bar{\mathcal{Z}} \subseteq \mathcal{Z}$ and $\bar{\mathcal{U}} \subseteq \mathcal{U}$. 
These tightened constraints are used to ensure robust constraint satisfaction and are defined in Section \ref{sec:closedloopperf}.
The solution of the OCP \eqref{eq:rompc} yields the optimized trajectory $(\bar{\mathbf{x}}^*_k, \bar{\mathbf{u}}^*_k)$, from which the simulated ROM control input at time $k$ is chosen as $\bar{u}_k = \bar{u}^*_{k|k}$. Since the next simulated ROM state $\bar{x}_{k+1}$ is computed via \eqref{eq:rom} with input $\bar{u}_k$, it also holds that $\bar{x}_{k+1} = \bar{x}^*_{k+1|k}$.

The positive definite matrix $P$ and the set $\bar{\mathcal{X}}_f$ define a terminal cost and terminal constraint that are designed to guarantee that the simulated nominal ROM with control $\bar{u}_k = \bar{u}^*_k$ is exponentially stable such that $\bar{x}_k \rightarrow 0$ and $\bar{u}_k \rightarrow 0$, and that the OCP \eqref{eq:rompc} is recursively feasible. 
Since methods for choosing $P$ and $\bar{\mathcal{X}}_f$ to achieve exponential stability for nominal full-state feedback MPC have been well established \cite[Chapter~2]{RawlingsMayneEtAl2017}, they are not discussed in detail here and instead we focus on leveraging those results in the context of ROMPC. Briefly, we note that in this work we compute $P$ and $\bar{\mathcal{X}}_f$ as outlined in detail in \cite[Section~IV]{LorenzettiLandryEtAl2019}, using the method from \cite{KolmanovskyGilbert1998} for computing positive invariant sets.
For exponential stability we also note that the following assumption is required.
\begin{ass}
$Q$ and $R$ are symmetric, positive definite.
\end{ass}
In practice, the matrix $Q$ in \eqref{eq:romcost} may not be positive definite. In this case, the state cost in \eqref{eq:romcost} can be made positive definite by using $Q + \gamma I$, where $\gamma \in \mathbb{R}$ is a small positive value.

\subsection{ROMPC Algorithm} \label{subsec:rompcalg}
The offline and online components of the ROMPC algorithm are now summarized, beginning with the offline controller synthesis procedure (Algorithm \ref{alg:offline}). The approach is decomposed such that the computationally intensive components are performed in the offline phase, which includes the computation of error bounds $\Delta_z$ and $\Delta_u$ that are used to define the tightened constraints $\bar{\mathcal{Z}}$ and $\bar{\mathcal{U}}$.

\begin{algorithm}
\caption{ROMPC Synthesis (Offline)}\label{alg:offline}
\begin{algorithmic}[1]
\Procedure{ROMPCSynthesis}{}
\State Compute $K$, $L$ (Section \ref{sec:controllersynth}, Algorithm \ref{alg:controllersynth})
\State Compute $\Delta_z$, $\Delta_u$ (Section \ref{sec:errorbounds}, Algorithm \ref{alg:ebounds})
\State Compute $\bar{\mathcal{Z}}$, $\bar{\mathcal{U}}$ (Section \ref{sec:closedloopperf}, Equation \eqref{eq:tightconstraints})
\State Compute $P$, $\bar{\mathcal{X}}_f$ (Section \ref{subsec:roocp})
\EndProcedure
\end{algorithmic}
\end{algorithm}

In the online portion of the ROMPC scheme (Algorithm \ref{alg:online}) the simulated ROM state $\bar{x}_k$ and reduced order state estimate $\hat{x}_k$ are initialized at time $k=0$ to be $\bar{x}_0 = \hat{x}_0 = 0$, and it is assumed that the ROMPC controller takes over at time $k=k_0$. 
For times $0 \leq k < k_0$ it is assumed that a ``startup'' controller is used to give the state estimator and simulated ROM time to converge from their initialized values toward values with smaller error, which is formalized in Assumption \ref{ass:epsilon0}.
In practice, the startup controller could be a simple linear feedback controller designed around a conservative (with respect to any state constraints), sub-optimal steady state operating point, or even a simple open-loop sequence. At time $k=k_0$ the ROMPC controller takes over to drive the system to different operating points for better performance, while guaranteeing constraint satisfaction.
It is also assumed that at time $k_0$ the OCP \eqref{eq:rompc} is feasible.
\begin{algorithm}
\caption{ROMPC Control (Online)}\label{alg:online}
\begin{algorithmic}[1]
\Procedure{ROMPC}{}
\State $k \leftarrow 0$, $\bar{x}_k \leftarrow 0$, $\hat{x}_k \leftarrow 0$
\Loop
\If{$k \geq k_0$}
\State $\bar{u}_k \leftarrow$ solveOCP($\bar{x}_k$)
\State $u_k \leftarrow \bar{u}_k + K(\hat{x}_k - \bar{x}_k)$
\Else
\State $u_k \leftarrow$ startupControl()
\State $\bar{u}_k \leftarrow u_k - K(\hat{x}_k - \bar{x}_k)$
\EndIf
\State $y_k \leftarrow$ getMeasurement()
\State applyControl($u_k$)
\State $\bar{x}_{k+1} \leftarrow A\bar{x}_k + B\bar{u}_k$
\State $\hat{x}_{k+1} \leftarrow A\hat{x}_k + Bu_k + L(y_k - C\hat{x}_k)$
\State $k \leftarrow k + 1$
\EndLoop
\EndProcedure
\end{algorithmic}
\end{algorithm}

\begin{figure*}[ht]
\centering
\includegraphics[width=0.7\textwidth]{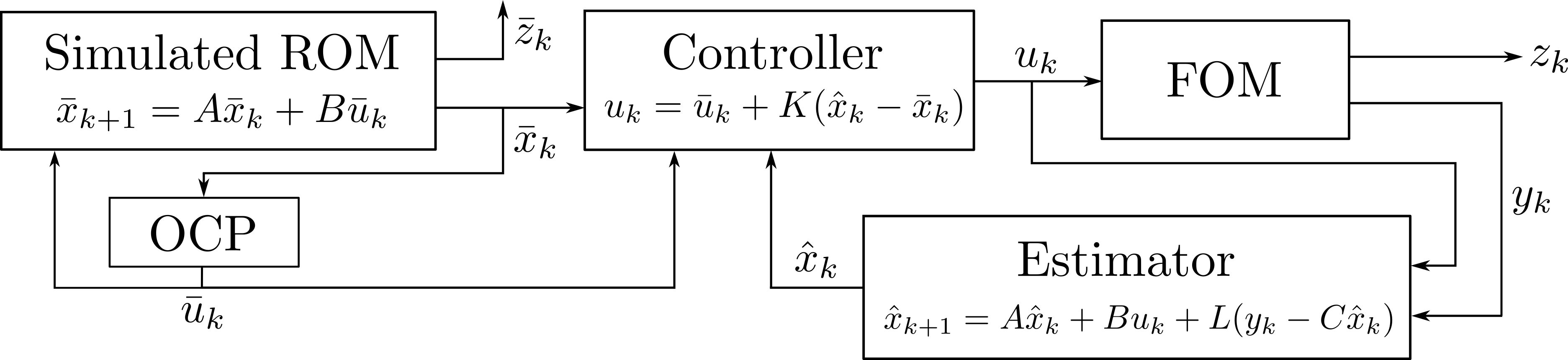}
\caption{A block diagram of the ROMPC control scheme, which shows the connection between the simulated ROM and the controlled system (FOM). The optimal control problem is used to control the simulated ROM, and the system (FOM) is driven by the controller to track the simulated ROM.}
\label{fig:blockdiagram}
\end{figure*}

For additional clarity, the architecture of the ROMPC control scheme (for times $k \geq k_0$) is shown in Figure \ref{fig:blockdiagram}. This diagram highlights the fact that there are two systems being controlled: the FOM \eqref{eq:fom} is the physical plant, and the ROM is \textit{simulated}.
The two systems are connected only by the controller \eqref{eq:controller}, which is essentially trying to drive the FOM to track the trajectory of the simulated ROM. 

Additionally, when $k < k_0$ and the startup controller is applied to the FOM, the simulated ROM is controlled by 
\begin{equation*}
    \bar{u}_k = u_k - K(\hat{x}_k - \bar{x}_k).
\end{equation*}
In this case the controller has essentially been reversed, and the simulated ROM tracks the trajectory of the FOM. This ensures a good initial condition $\bar{x}_{k_0}$ is used when the OCP takes control of the simulated ROM. This approach for initializing the ROM state $\bar{x}_{k_0}$ is used rather than just setting $\bar{x}_{k_0} = \hat{x}_{k_0}$ because the history $\bar{x}_{k}$ for $k < k_0$ is used in the computation of the error bounds in Section \ref{sec:errorbounds}.

\section{Closed-Loop Performance} \label{sec:closedloopperf}
This section\footnote{While we primarily leverage existing techniques from MPC, we include this section for completeness and clarity of the overall approach.} discusses two key properties of the closed-loop systems' performance, namely robust constraint satisfaction and stability. In the context of this work, a property is considered \textit{robust} if it holds in the presence of ROM approximation errors, state estimation errors, and admissible bounded disturbances \eqref{eq:noise}.

\begin{figure}[ht]
\centering
\includegraphics[width=0.9\columnwidth]{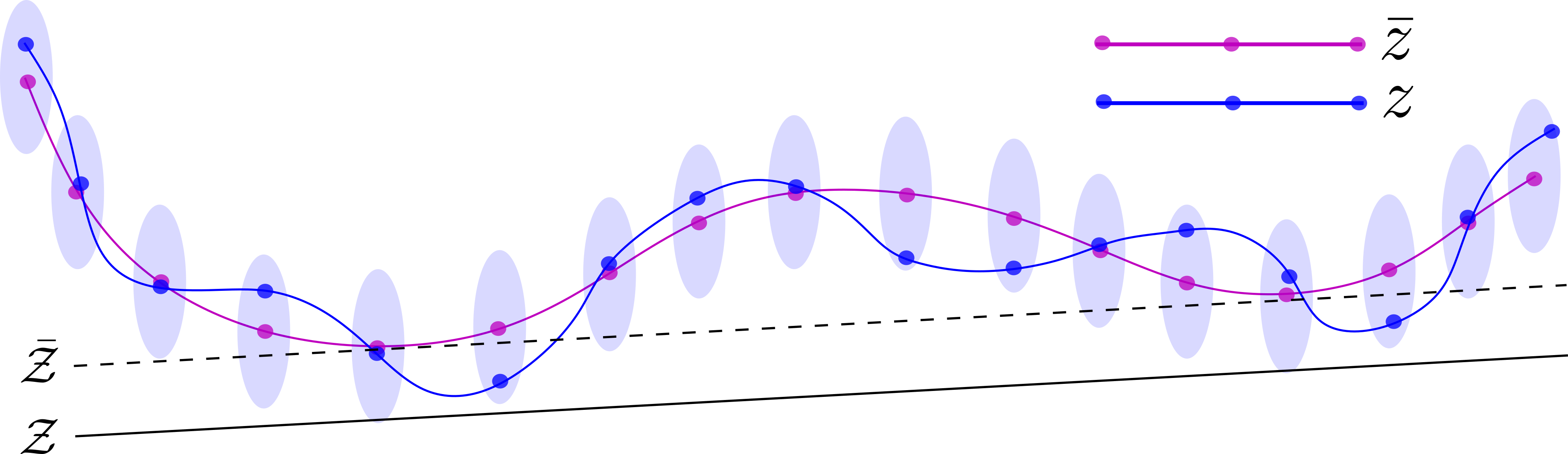}
\caption{Robust constraint satisfaction can be guaranteed by using error tubes and constraint tightening. The trajectory of the simulated ROM \eqref{eq:rom}, which is controlled by the OCP \eqref{eq:rompc} and satisfies the tightened constraints $\bar{z} \in \bar{\mathcal{Z}}$, is shown in pink. The performance variables, $z$, of the full order system \eqref{eq:fom} are shown in blue, and track $\bar{z}$ with bounded error. The tightened constraint $\bar{\mathcal{Z}}$ guarantees that the entire tube, and thus $z$, satisfies the constraint $z \in \mathcal{Z}$.}
\label{fig:tubes}
\vspace{-8pt}
\end{figure}

Recall that the reduced order OCP \eqref{eq:rompc} optimizes the trajectory of the simulated ROM under the constraints that $\bar{z} \in \bar{\mathcal{Z}}$ and $\bar{u} \in \bar{\mathcal{U}}$. The control law \eqref{eq:controller} and state estimator \eqref{eq:estimator} then make the FOM \eqref{eq:fom} track this trajectory.
However, perfect tracking is not possible due to disturbances, model reduction errors, and state estimation errors. Thus, choosing $\bar{\mathcal{Z}} = \mathcal{Z}$ and $\bar{\mathcal{U}} = \mathcal{U}$ will not \textit{guarantee} constraint satisfaction for the FOM.
A tube-based approach \cite[Chapter~3.5]{RawlingsMayneEtAl2017} is therefore used to robustly guarantee constraint satisfaction, where the worst-case tracking errors are quantified (in Section \ref{sec:errorbounds}) and used to define ``tubes'' around the nominal trajectories of $\bar{z}$ and $\bar{u}$. Since these tubes are guaranteed to contain the actual system trajectories, requiring that the entire tube satisfy the constraints will guarantee the actual trajectory will also satisfy the constraints.
In practice, this is implemented by choosing $\bar{\mathcal{Z}}$ and $\bar{\mathcal{U}}$ to be \textit{tightened} versions of the original constraints, as represented visually in Figure \ref{fig:tubes}.

\subsection{Closed-Loop Error Dynamics}
The worst-case errors $\delta_z \coloneqq z - \bar{z}$ and $\delta_u \coloneqq u - \bar{u}$ are quantified by analyzing the closed-loop error dynamics of the joint error state $\epsilon \coloneqq [e^T,\: d^T]^T$, with state reduction error $e \coloneqq x^f - V\bar{x}$ and control error $d = \hat{x} - \bar{x}$, given by
\begin{equation} \label{eq:edynamics}
\begin{split}
\epsilon_{k+1} &= A_\epsilon \epsilon_k + B_\epsilon r_k + G_\epsilon \omega_k, \\
\end{split}
\end{equation}
where
\begin{equation*}
\begin{split}
A_\epsilon &= \begin{bmatrix}
A^f & B^fK \\ LC^f & A + BK - LC
\end{bmatrix}, \\
B_\epsilon &= \begin{bmatrix}
P_\perp A^fV & P_\perp B^f \\ 0 & 0
\end{bmatrix}, \quad G_\epsilon = \begin{bmatrix}
B^f_w & 0 \\ 0 & L
\end{bmatrix},
\end{split}
\end{equation*}
and with $P_\perp = I-V(W^TV)^{-1}W^T$, $r = [\bar{x}^T, \: \bar{u}^T]^T$, and $\omega = [w^T,\:v^T]^T$.
Note that $\delta_z = H^f e$ and $\delta_u = Kd$ can be considered outputs of this error system, and the inputs are the disturbances, $(w,v)$, and the simulated ROM trajectory, $(\bar{x},\bar{u})$.

\subsection{Closed-Loop Robust Constraint Satisfaction}
The closed-loop error dynamics \eqref{eq:edynamics} provide a means for analyzing the worst-case tracking errors $\delta_z$ and $\delta_u$. Of particular interest is how the tracking errors could lead to constraint violations. Since the constraint sets $\mathcal{Z}$ and $\mathcal{U}$ defined in \eqref{eq:constraints} are polytopes expressed by a set of half-spaces, from the definition of the state reduction error $e$ and control error $d$ it follows that
\begin{equation*}
\begin{split}
z \in \mathcal{Z} &\iff H_z(\bar{z} + \delta_z)  \leq b_z, \\
u \in \mathcal{U} &\iff H_u(\bar{u} + \delta_u)  \leq b_u.
\end{split}
\end{equation*}
With a set of worst-case bounds $\Delta_z \in \mathbb{R}^{n_z}$ and $\Delta_u \in \mathbb{R}^{n_u}$ on the tracking error, such that $H_z \delta_z \leq \Delta_z$ and $H_u \delta_u \leq \Delta_u$, the OCP constraint sets $\bar{\mathcal{Z}}$, $\bar{\mathcal{U}}$ can be defined as 
\begin{equation} \label{eq:tightconstraints}
\begin{split}
\bar{\mathcal{Z}} \coloneqq \{ \bar{z} \:|\: H_z \bar{z} \leq b_z - \Delta_z \}, \\
\bar{\mathcal{U}} \coloneqq \{ \bar{u} \:|\: H_u \bar{u} \leq b_u - \Delta_u \}, \\
\end{split}
\end{equation}
which are tightened versions of $\mathcal{Z}$ and $\mathcal{U}$. 
With these tightened constraint sets, the nominal trajectory computed by the OCP \eqref{eq:rompc} will account for the tracking error and therefore ensure constraint satisfaction. An efficient method for computing the bounds $\Delta_z$ and $\Delta_u$ is presented in Section \ref{sec:errorbounds}.
\begin{lem}[Robust Constraint Satisfaction] \label{lem:robustconstraint}
Suppose that at time $k_0$ the optimal control problem \eqref{eq:rompc} is feasible and that $H_z\delta_{z,k} \leq \Delta_z$ and $H_u \delta_{u,k} \leq \Delta_u$ for all $k \geq k_0$. Then, under the proposed control scheme the full order system will robustly satisfy the constraints \eqref{eq:constraints} for all $k\geq k_0$.
\end{lem}

\subsection{Closed-Loop Stability}
In addition to robustly satisfying the constraints, the full order system \eqref{eq:fom} must be stable.
In the disturbance free case (i.e. when $w = 0$ and $v = 0$), the ROMPC scheme defined in Section \ref{sec:rompc} is guaranteed to make \eqref{eq:fom} stable and converge asymptotically to the origin. However, guaranteed convergence to the origin is not possible when unpredictable disturbances affect the system. But, if the disturbances $w$ and $v$ are \textit{bounded}, the controlled full order system can be shown to asymptotically converge to a compact set containing the origin (whose size depends on the disturbance bounds) under certain assumptions.\footnote{Computation of this set is theoretically possible \cite{KolmanovskyGilbert1998}, but is computationally intractable for high-dimensional systems and often not practically useful.}

\begin{thm}[Robust Stability] \label{thm:robuststability}
Suppose that $A_\epsilon$ is Schur stable, that the OCP \eqref{eq:rompc} is feasible at some time $k_0$, and that the OCP drives the simulated ROM trajectory to the origin exponentially fast such that $\lVert \bar{x}_k\rVert \leq M\gamma^k$ for all $k \geq k_0$ and for some values $M > 0$ and $\gamma \in (0,1)$.
Then, the closed-loop system robustly and  asymptotically converges to a compact set containing the origin.
\end{thm}

\begin{crl}[Stability] \label{crl:stability}
Suppose the conditions from Theorem \ref{thm:robuststability} hold.
Then, in the disturbance free case where $\mathcal{W} =\{0\}$ and $\mathcal{V} = \{0\}$, the origin is asymptotically stable for the closed-loop system.
\end{crl}

The requirement that $A_\epsilon$ is Schur stable is natural, and is satisfied by properly choosing the gain matrices $K$ and $L$. In fact, without the model order reduction step (i.e. $A = A^f$, $B = B^f$, $C = C^f$) the separation principle guarantees the stability of the error dynamics by the Schur stability of $A + BK$ and $A-LC$. Unfortunately, with the use of a reduced order model the separation principle cannot be applied since $A^f \neq A$, and therefore no such guarantee on the stability of $A_\epsilon$ can be made. A methodology for synthesizing the controller gains $K$ and $L$ such that $A_\epsilon$ is Schur stable is presented in Section \ref{sec:controllersynth}.

\section{Controller Synthesis} \label{sec:controllersynth}
In this section a methodology for synthesizing the gain matrices $K$ and $L$ is presented. The goal of the synthesis method is to find a set of gains that will ensure that the error dynamics are stable (i.e. $A_\epsilon$ is Schur stable), and that the tracking errors $\delta_z$ and $\delta_u$ remain small. 

In principle, stability is straightforward to verify since $A_\epsilon$ is Schur stable if all of its eigenvalues satisfy $\lvert \lambda \rvert < 1$. 
Additionally, the metric chosen to benchmark the tracking performance is the $\mathcal{H}_2$ system norm of the closed-loop error dynamics \eqref{eq:edynamics} with outputs
\begin{equation}
\tilde{z}_k = H_\epsilon \epsilon_k, \quad H_\epsilon = \begin{bmatrix}W_z H^f & 0 \\ 0 & W_u K \end{bmatrix},
\end{equation}
where $W_z$ and $W_u$ are user defined weighting matrices.
Note that the outputs $\tilde{z} = [(W_z\delta_z)^T,\: (W_u\delta_u)^T]^T$ are the weighted tracking errors and the inputs are the combined model reduction errors and disturbances, which can be concatenated $\tilde{w} \coloneqq [r^T,\omega^T]^T$. 

The $\mathcal{H}_2$-norm of a closed-loop system with inputs $\tilde{w}$ and output $\tilde{z}$, denoted by $\lVert G \rVert_{\mathcal{H}_2}$ (where $G$ is the closed-loop transfer function from inputs $\tilde{w}$ to outputs $\tilde{z}$), satisfies the property
\begin{equation*}
\lVert \tilde{z} \rVert_{L_\infty} \leq \lVert G \rVert_{\mathcal{H}_2} \lVert \tilde{w} \rVert_{L_2}
\end{equation*}
where $\lVert \tilde{z} \rVert_{L_\infty} = \sup_{k \geq 0} \lVert z_k \rVert_\infty$ and $ \lVert \tilde{w} \rVert_{L_2} = \sqrt{\sum_{k=0}^\infty \lVert \tilde{w}_k \rVert_2^2}$ \cite{BennerGugercinEtAl2015}. In other words, the $\mathcal{H}_2$-norm of the closed-loop system provides an upper bound on the maximum amplitude of the output $\tilde{z}$ for inputs $\tilde{w}$ of bounded energy. Since the outputs $\tilde{z}$ in this problem correspond to the tracking errors $\delta_z$ and $\delta_u$, minimizing the $\mathcal{H}_2$-norm can be used as a surrogate for minimizing the worst-case tracking error. Controller synthesis for minimizing the closed-loop $\mathcal{H}_2$-norm is commonly referred to as $\mathcal{H}_2$-optimal control \cite{ZhouDoyleEtAl1996, Petersson2013}.

\subsection{\texorpdfstring{$\mathcal{H}_2$}{H2}-Optimal Controller Synthesis}
The $\mathcal{H}_2$-optimal controller synthesis problem is now written in a more standard form.
First, the closed-loop dynamics \eqref{eq:edynamics} are algebraically rearranged into the \textit{equivalent} system
\begin{equation} \label{eq:h2errordynamics}
\begin{split}
\epsilon_{k+1} &= \tilde{A}\epsilon_k + \tilde{B} \tilde{u}_k + \tilde{B}_{\tilde{w}} \tilde{w}_k, \\
\tilde{y}_k &= \tilde{C} \epsilon_k + \tilde{D}_{\tilde{y}\tilde{w}} \tilde{w}_k, \\
\tilde{z}_k &= \tilde{H} \epsilon_k + \tilde{D}_{\tilde{z}\tilde{u}} \tilde{u}_k,\\
\tilde{u}_k &= \tilde{K} \tilde{y}_k,
\end{split}
\end{equation}
where
\begin{equation*}
\begin{split}
&\tilde{A} \coloneqq \begin{bmatrix}
A^f & 0 \\ 0 & A
\end{bmatrix}, \quad \tilde{B} \coloneqq \begin{bmatrix}
0 & B^f \\ I & B
\end{bmatrix}, \quad \tilde{C} \coloneqq \begin{bmatrix}
C^f & -C \\ 0 & I
\end{bmatrix}, \\
&\tilde{H} \coloneqq \begin{bmatrix}
W_z H^f & 0 \\ 0 & 0
\end{bmatrix}, \quad \tilde{K} \coloneqq \begin{bmatrix}
L & 0 \\ 0 & K
\end{bmatrix},
\end{split}
\end{equation*}
and with noise matrices
\begin{equation*}
\begin{split}
&\tilde{B}_{\tilde{w}} \coloneqq \begin{bmatrix}
P_\perp A^fV & P_\perp B^f & B^f_w & 0 \\ 0 & 0 & 0 & 0
\end{bmatrix}, \\ 
&\tilde{D}_{\tilde{y}\tilde{w}} \coloneqq \begin{bmatrix}
0 & 0 & 0 & I \\ 0 & 0 & 0 & 0
\end{bmatrix}, \quad \tilde{D}_{\tilde{z}\tilde{u}} \coloneqq \begin{bmatrix}
0 & 0 \\ 0 & W_u
\end{bmatrix}.
\end{split}
\end{equation*}

With the requirement that $\tilde{u} = \tilde{K}\tilde{y}$, it can be seen that the problem of choosing $K$ and $L$ to minimize the $\mathcal{H}_2$-norm of the closed-loop error dynamics can be cast as solving an $\mathcal{H}_2$-optimal control problem with a structured, static output feedback controller. This problem is compactly expressed as
\begin{equation} \label{eq:fullH2opt}
\min_{K,L} \:\: \lVert G \rVert_{\mathcal{H}_2},
\end{equation}
where $G$ is the closed-loop transfer function of \eqref{eq:h2errordynamics} (or equivalently of \eqref{eq:edynamics}) with inputs $\tilde{w}$ and outputs $\tilde{z}$.

There exist several approaches for solving such problems, including optimization methods based on convex approximations or non-convex optimization \cite{SadabadiPeaucelle2016}. However, for extremely high-dimensional problems these approaches may not be computationally tractable. One alternative to address the computational issues arising from high-dimensional problems is to approximate the objective function by a surrogate that is more efficient to evaluate. This approach is used for $\mathcal{H}_\infty$-optimal controller synthesis in \cite{MitchellOverton2015} and \cite{BennerMitchellEtAl2018}. Of particular interest, \cite{BennerMitchellEtAl2018} minimizes the $\mathcal{H}_\infty$ norm of a reduced order system instead of the original high-dimensional one.

We propose to use a similar approach to \cite{BennerMitchellEtAl2018}: approximately solve the original problem by minimizing the $\mathcal{H}_2$-norm of a \textit{reduced order} error system that approximates \eqref{eq:h2errordynamics}.
Consider a reduced order approximation defined by
\begin{equation} \label{eq:h2reduceddynamics}
\begin{split}
\epsilon_{r,k+1} &= \tilde{A}_r\epsilon_{r,k} + \tilde{B}_r \tilde{u}_k + \tilde{B}_{\tilde{w},r} \tilde{w}_k, \\
\tilde{y}_{r,k} &= \tilde{C}_r \epsilon_{r,k} + \tilde{D}_{\tilde{y}\tilde{w}}\tilde{w}_k, \\
\tilde{z}_{r,k} &= \tilde{H}_{r} \epsilon_{r,k} + \tilde{D}_{\tilde{z}\tilde{u}}\tilde{u}_k,\\
\tilde{u}_k &= \tilde{K}\tilde{y}_k, \\
\end{split}
\end{equation}
where $\tilde{A}_r = (\tilde{W}^T\tilde{V})^{-1}\tilde{W}^T \tilde{A}\tilde{V}$, $\tilde{B}_r = (\tilde{W}^T\tilde{V})^{-1}\tilde{W}^T \tilde{B}$, $\tilde{B}_{\tilde{w},r} = (\tilde{W}^T\tilde{V})^{-1}\tilde{W}^T\tilde{B}_{\tilde{w}}$, $\tilde{C}_{r} = \tilde{C}\tilde{V}$, $\tilde{H}_{r} = \tilde{H}\tilde{V}$, and $\tilde{V}$ and $\tilde{W}$ are the reduced order basis that define a Petrov-Galerkin projection. The approximate $\mathcal{H}_2$-optimal control problem is
\begin{equation} \label{eq:reducedH2opt}
\min_{K,L} \:\: \lVert G_r \rVert_{\mathcal{H}_2},
\end{equation}
where $G_r$ is the closed-loop transfer function of the reduced order closed-loop error dynamics system \eqref{eq:h2reduceddynamics}. This problem is much more computationally tractable and is generally amenable to the previously mentioned techniques for computing structured, static output feedback controllers. Additionally, the basis matrices used to define the ROM \eqref{eq:orig_rom} can be directly reused in \eqref{eq:h2reduceddynamics} rather than performing an additional model reduction procedure. In Section \ref{subsec:rorm} we show that this leads to further simplifications of the control synthesis problem.

\subsection{Reduced Order Riccati Method} \label{subsec:rorm}
By reusing the basis matrices $V$ and $W$ that define the ROM \eqref{eq:orig_rom} to define \eqref{eq:h2reduceddynamics}, the problem \eqref{eq:reducedH2opt} reduces to a classical $\mathcal{H}_2$-optimal control problem that only requires the solution of two Riccati equations.
Specifically, the matrices $\tilde{V}$ and $\tilde{W}$ are chosen as
\begin{equation*}
\tilde{V} = \begin{bmatrix}
V & 0 \\ 0 & I
\end{bmatrix}, \quad \tilde{W} = \begin{bmatrix}
W & 0 \\ 0 & I
\end{bmatrix}.
\end{equation*}
In this case the reduced order closed-loop error system \eqref{eq:h2reduceddynamics} can be algebraically re-written as
\begin{equation} \label{eq:h2ricattimethod}
\begin{split}
e_{r,k+1} &= Ae_{r,k} + Bu_{r,k} + \begin{bmatrix} B_w & 0 \end{bmatrix} \omega_k, \\
y_{r,k} &= Ce_{r,k} + \begin{bmatrix}
0 & I
\end{bmatrix}\omega_k, \\
\tilde{z}_{r,k} &= \begin{bmatrix}
W_zH  \\ 0 
\end{bmatrix}e_{r,k} + \begin{bmatrix}
 0 \\  W_u
\end{bmatrix}u_{r,k}, \\
d_{k+1} &= Ad_k + Bu_{r,k} + L(y_{r,k} -Cd_k), \\
u_{r,k} &= Kd_k, \\
\end{split}
\end{equation}
where $e_r$ is the reduced order approximation of the state reduction error $e$. 

There are two interesting things to note about these dynamics. First, they are exactly the dynamics that would be used for controller synthesis in standard problems where there is no model order reduction and only the reduced order system was considered (i.e. $A^f = A$, $B^f = B$, $C^f = C$, $H^f=H$). In other words, this is a classical $\mathcal{H}_2$-optimal controller synthesis problem applied to the ROM. 
Second, for the simplified choice of $\tilde{V}$ and $\tilde{W}$ the solution to the optimization problem \eqref{eq:reducedH2opt} is
\begin{equation}
\begin{split}
K = -(B^TXB + R)^{-1}B^TXA, \\
L = AYC^T(CYC^T + I)^{-T},
\end{split}
\end{equation}
where $X$ and $Y$ are solutions of the discrete-time algebraic Riccati equations
\begin{equation} \label{eq:h2riccati}
\begin{split}
X &= A^TXA - A^TXB(B^TXB + R)^{-1}B^TXA + Q_z, \\
Y &= AYA^T - AYC^T(CYC^T + I)^{-1}CYA^T + Q_w, \\
\end{split}
\end{equation}
where $R = W_u^TW_u$, $Q_z = H^TW_z^TW_zH$, $Q_w = B_wB_w^T + \gamma I$ with $\gamma > 0$, and $R$ is assumed to be full rank.
Note that the identity matrix in the choice of $Q_w$ is a regularization term that is added to indirectly account for the model reduction disturbances. This is required because the use of $V$ and $W$ to simplify the $\mathcal{H}_2$-optimal synthesis problem leads to the model reduction disturbance terms to be completely ignored in \eqref{eq:h2ricattimethod}.

The reduced order Riccati method is summarized in Algorithm \ref{alg:controllersynth}. Note that no guarantees on the stability of $A_\epsilon$ are provided since the reduced order system is used to approximate the true problem. Therefore it is imperative to check the stability \textit{a posteriori}.\footnote{Verifying the stability of $A_\epsilon$ can be efficient even if the full order system is high-dimensional, assuming it is sparse. In particular, only the largest magnitude eigenvalue needs to be identified.} Options to consider if Algorithm \ref{alg:controllersynth} fails to stabilize $A_\epsilon$ are discussed in Section \ref{subsec:ctrlsynthremarks}.

\begin{algorithm}
\caption{Controller Gain Synthesis}\label{alg:controllersynth}
\begin{algorithmic}[1]
\Procedure{RedOrderRiccatiMethod}{...}
\State $R \leftarrow W_u^TW_u$ ($R$ full rank)
\State $Q_z \leftarrow H^TW_z^TW_zH$
\State $Q_w \leftarrow B_wB_w^T + \gamma I$
\State $X,Y \leftarrow$ solve Riccati equations \eqref{eq:h2riccati}
\State $K = -(B^TXB + R)^{-1}B^TXA$
\State $L = AYC^T(CYC^T + I)^{-T}$
\If{$A_\epsilon$ is Schur stable}{ \Return $K$, $L$}
\EndIf
\EndProcedure
\end{algorithmic}
\end{algorithm}

\subsection{Controller Synthesis Results}
The reduced order Riccati method (Algorithm \ref{alg:controllersynth}) solves the approximate $\mathcal{H}_2$-optimal control problem \eqref{eq:reducedH2opt} as a surrogate for the full order problem \eqref{eq:fullH2opt}. To benchmark the effectiveness of this method it is compared against an alternative technique that attempts to directly solve the original full order $\mathcal{H}_2$-optimal control problem \eqref{eq:fullH2opt}.
In particular, a quasi-Newton BFGS optimization method implemented in the package HIFOO \cite{ArzelierDeaconuEtAl2011} is used, which first minimizes the spectral radius until a stabilizing controller is found, and then minimizes the $\mathcal{H}_2$-norm. Since the $\mathcal{H}_2$-norm is non-convex and convergence to a global minimum is not guaranteed, HIFOO uses multiple random initial conditions and returns the best solution.

For the comparison, HIFOO is first used without any initial solution guess, it is then applied a second time with a warm-start by the solution from Algorithm \ref{alg:controllersynth}. Results for each of the three methods (Riccati, HIFOO, and HIFOO with Riccati warm-start) are presented in Table \ref{tab:ctrlsynth}, with the resulting computation times presented in Table \ref{tab:ctrlsynthtimes}. Several example problems are considered, which are described in Section \ref{sec:experiments}.\footnote{A modified version of the open-source HIFOO code is included in the repository accompanying this work, which is simplified to only solve the controller synthesis problem considered in this work and modified to handle discrete-time systems.} In Table \ref{tab:ctrlsynth} the values $R_{\mathcal{H}_2}$ are defined as
\begin{equation}
R_{\mathcal{H}_2} \coloneqq \frac{\lVert \Sigma - \Sigma_r \rVert_{\mathcal{H}_2}}{\lVert \Sigma \rVert_{\mathcal{H}_2}},
\end{equation}
where $\Sigma$ is the transfer function from $u$ to $z$ of the full order model \eqref{eq:fom} and $\Sigma_r$ is the transfer function from $\bar{u}$ to $\bar{z}$ of the reduced order model \eqref{eq:rom}. This value can therefore be used as a metric for how good the reduced order model approximates the true model with respect to the $\mathcal{H}_2$-norm, where a small value of $R_{\mathcal{H}_2}$ indicates a good approximation. The remaining values in Table \ref{tab:ctrlsynth} are the values of the $\mathcal{H}_2$-norm, $\lVert G \rVert_{\mathcal{H}_2}$, of the full order closed-loop error system \eqref{eq:h2errordynamics} defined with the synthesized gains $K$ and $L$ using either HIFOO or HIFOO with Riccati warm-start, and expressed as a fraction of the $\mathcal{H}_2$-norm resulting from the reduced order Riccati method.

These results show that the reduced order Riccati method performs comparably to direct optimization over a variety of problems. Additionally, from the computation times listed in Table \ref{tab:ctrlsynthtimes} it can be seen that the reduced order Riccati method drastically outperforms in efficiency. In general, the Riccati method only depends on the ROM dimension\footnote{The ROM dimension is a user-selected value, typically chosen with the aid of heuristics that are useful in the context of the specific model reduction method for quantifying model reduction error.} and does not depend on the dimension of the full order model, which makes it scalable to even extremely high-dimensional ROMPC applications assuming $n \ll n^f$.

\begin{table}[htbp]
\caption{Controller Synthesis Results}
\centering
\setlength\tabcolsep{5pt}
\renewcommand{\arraystretch}{1.2}
\begin{tabular}{|c|c|c|c|c|c|c|c|}
\cline{1-7}
\multicolumn{1}{|r|}{Model} & $n^f$ & $n$ & $m$ & $R_{\mathcal{H}_2}$ & HIFOO & HIFOO+Riccati \\ \hline
\multicolumn{1}{|r|}{Small Synthetic} & 6 & 4 & 1 & 0.057 & 0.999 & 1.000 \\ \hline
\multicolumn{1}{|r|}{Large Synthetic} & 20 & 6  & 1 & 0.009 & 0.935 & 0.950 \\ \hline
\multicolumn{1}{|r|}{Dist. Column} & 86 & 12 & 4 & 0.005 & 0.991 & 0.973 \\ \hline
\multicolumn{1}{|r|}{Tubular Reactor} & 600 & 30 & 3 & 0.016 & 0.999 & 1.000 \\ \hline
\multicolumn{1}{|r|}{Heatflow} & 3,481 & 21 & 5 & 0.087 & 1.003 & 0.995 \\ \hline
\multicolumn{7}{p{240pt}}{Results comparing the reduced order Riccati method (Algorithm \ref{alg:controllersynth} with $\gamma=0.001$) against a quasi-Newton approach applied directly to minimizing \eqref{eq:fullH2opt} (HIFOO), as well as a combined method where the Riccati method is used to warmstart HIFOO (HIFOO+Riccati). The values $R_{\mathcal{H}_2} = \lVert \Sigma - \Sigma_r \rVert_{\mathcal{H}_2}/\lVert \Sigma \rVert_{\mathcal{H}_2}$ are used to quantify the model reduction error. The values corresponding to the different methods are the $\mathcal{H}_2$-norms of the closed-loop error system, $\lVert G \rVert_{\mathcal{H}_2}$, expressed as a ratio corresponding to the value computed using the reduced order Riccati method. For example a value of 0.95 implies a 5\% smaller value of $\lVert G \rVert_{\mathcal{H}_2}$ with respect to the Riccati method (Algorithm \ref{alg:controllersynth}).}\\
\end{tabular}
\label{tab:ctrlsynth}
\vspace{-10pt}
\end{table}

\begin{table}[htbp]
\caption{Controller Synthesis Computation Times}
\centering
\setlength\tabcolsep{5pt}
\renewcommand{\arraystretch}{1.2}
\begin{tabular}{|c|c|c|c|c|c|}
\cline{1-6}
\multicolumn{1}{|r|}{Model} & $n^f$ & $n$ & Riccati & HIFOO & HIFOO+Riccati \\ \hline
\multicolumn{1}{|r|}{Small Synthetic} & 6 & 4 & 9 ms & 422 ms & 244 ms \\ \hline
\multicolumn{1}{|r|}{Large Synthetic} & 20 & 6 & 3 ms & 544 ms & 229 ms \\ \hline
\multicolumn{1}{|r|}{Dist. Column} & 86 & 12 & 8 ms & 64 s & 17 s \\ \hline
\multicolumn{1}{|r|}{Tubular Reactor} & 600 & 30 & 24 ms & 683 s & 12 s \\ \hline
\multicolumn{1}{|r|}{Heatflow} & 3,481 & 21 & 18 ms & 47.5 h & 6.9 h \\ \hline
\multicolumn{6}{p{240pt}}{Computation times for synthesizing the controll gains $K$ and $L$ using the reduced order Riccati method (Riccati), a quasi-Newton approach applied directly to minimizing \eqref{eq:fullH2opt} (HIFOO), as well as a combined method where the Riccati method is used to warmstart HIFOO (HIFOO+Riccati). The reduced order Riccati method is much faster because it is only dependent on the dimension $n$ of the reduced order model.}\\
\end{tabular}
\label{tab:ctrlsynthtimes}
\vspace{-10pt}
\end{table}

\subsection{Remarks} \label{subsec:ctrlsynthremarks}
The primary disadvantage of the reduced order Riccati method is the lack of \textit{guarantees} that the resulting controller will result in the closed-loop system being asymptotically stable. While good performance has been empirically demonstrated, similar results may not hold if the reduced order models are not sufficiently good approximations. In the case that the reduced order Riccati method fails to stabilize the full order system, two additional options should be considered. First, the accuracy of the reduced order model should be verified (e.g. by simulation). This is a worthwhile first step because even if a set of stabilizing gains $K$ and $L$ could be identified with a different approach, poor model approximation could still lead to large tracking errors and degraded performance for the closed-loop system. Second, for moderately-sized problems an optimization problem that incorporates the full order closed-loop error system \eqref{eq:h2errordynamics} could be solved.

As can be seen from the previous results, directly optimizing the $\mathcal{H}_2$-norm of the full order closed-loop error system becomes challenging with high-dimensional problems. In particular, HIFOO becomes computationally challenging in these cases because each gradient evaluation of the $\mathcal{H}_2$-norm, $\lVert G \rVert_{\mathcal{H}_2}$, with respect to the controller gains $K$ and $L$ involves solving full order Lyapunov equations. A slightly more practical method, that would still guarantee stability of the closed-loop system, is to consider an optimization problem whose objective is the reduced order approximate $\mathcal{H}_2$-norm, $\lVert G_r \rVert_{\mathcal{H}_2}$, but is subjected to a stability constraint on the full order system. This is the approach proposed for reduced order $\mathcal{H}_\infty$ controller synthesis in \cite{BennerMitchellEtAl2018}. The advantage of this approach is that the full order system is only used to compute the spectral radius (i.e. to check stability) or compute a gradient of the spectral radius, and both of these operations can leverage efficient sparse eigenvalue methods.

\section{Error Bounds for Constraint Tightening} \label{sec:errorbounds}
This section introduces a methodology for computing bounds to quantify the worst-case tracking error between the FOM \eqref{eq:fom} and the simulated ROM \eqref{eq:rom}. In particular these error bounds, denoted by $\Delta_z$ and $\Delta_u$, are used to tighten the constraint sets \eqref{eq:tightconstraints} to guarantee robust constraint satisfaction.
The two primary challenges associated with such an analysis are computational efficiency and the amount of conservatism introduced. Computational efficiency is important since the error dynamics system \eqref{eq:edynamics} is high-dimensional, and conservatism is important because excessive constraint tightening can lead to sub-optimal performance. Unfortunately, since the error bounds are computed \textit{a priori} (i.e. consider worst-case scenarios), a certain amount of conservatism is unavoidable.

An approach from linear robust MPC theory is to compute a robust positively invariant set for the error dynamics, which is a set that is invariant under the error dynamics with \textit{any} admissible disturbance. However, not only do algorithms for computing these sets scale poorly with problem dimension, this approach considers bounded disturbances that are not correlated in time and therefore can not leverage the structure of the model approximation error (i.e. it is induced by the \textit{trajectory} of the simulated ROM). This could lead to an over-approximation of the error and thus be overly conservative.

Two approaches for error analysis in ROMPC have been previously proposed in \cite{LoehningRebleEtAl2014, KoegelFindeisen2015, LorenzettiLandryEtAl2019}. Linear programming is used in \cite{KoegelFindeisen2015, LorenzettiLandryEtAl2019}, but in both cases the full order dynamics model \eqref{eq:fom} is embedded in the constraints, and thus these approaches do not scale well with problem dimension. The approach used in \cite{LoehningRebleEtAl2014} produces a bound on the \textit{norm} of the error state and is slightly more computationally efficient, but at the cost of being more conservative as shown in \cite{LorenzettiPavone2020}. Similar error analyses using norms are discussed in \cite{HaasdonkOhlberger2011, HasenauerLoehningEtAl2012}. The approach used in this work can be viewed as a blend of the two approaches.

\subsection{Summary of Approach} \label{subsec:ebound_summary}
The error bounds $\Delta_z$ and $\Delta_u$ that are used to tighten the constraints in \eqref{eq:tightconstraints} are bounds on $H_z \delta_z$ and $H_u \delta_u$, which can also be written as $E_z \epsilon$ and $E_u \epsilon$ where
\begin{equation}
E_z = \begin{bmatrix}
H_z H^f & 0
\end{bmatrix}, \quad E_u = \begin{bmatrix}
 0 & H_u K
\end{bmatrix}.
\end{equation}
Therefore, to compute worst-case bounds for these quantities the trajectories of the error dynamics \eqref{eq:edynamics} must be analyzed. Instead of using the recursive form of the dynamics, consider the equivalent form
\begin{equation} \label{eq:error_response}
\epsilon_{k} = A_\epsilon^{k-\underline{k}} \epsilon_{\underline{k}} + \sum_{j=\underline{k}}^{k-1}A_\epsilon^{k-1-j}\Big( B_\epsilon r_j + G_\epsilon \omega_j\Big),
\end{equation}
for any time $k > \underline{k}$. Additionally, the time $\underline{k}$ is defined as $\underline{k} = k_0 - 2\tau$ where
$k_0$ is the time when the ROMPC scheme takes control of the full order system and $\tau$ is a user-defined time horizon (discussed at the end of Section \ref{subsec:ebound_summary}). From this representation of the error, it can be seen that $\epsilon_k$ is defined for all $k \geq \underline{k}$ by the error $\epsilon_{\underline{k}}$, the trajectory of the simulated ROM, $r_k$, and the disturbances, $\omega_k$. The disturbance terms have already been assumed to be bounded (Assumption \ref{ass:compact}), but the following additional assumptions on $\epsilon_{\underline{k}}$ and $r_k$ are also made to ensure that $\epsilon_k$ is bounded for all $k \geq \underline{k}$:

\begin{ass} \label{ass:inf_hor_lp}
The simulated ROM satisfies $\bar{u}_k \in \mathcal{U}$ and $\bar{z}_k\in\mathcal{Z}$ for all $k \geq \underline{k}$,
\end{ass}
\begin{ass} \label{ass:epsilon0}
The error $\epsilon_{\underline{k}}$ is bounded by $\lVert \epsilon_{\underline{k}} \rVert_G \leq \eta_{\underline{k}}$,
\end{ass}
where $G$ is some positive definite weighting matrix such that $\lVert \cdot \rVert_G = \sqrt{(\cdot)^TG(\cdot)}$.

Note that Assumption \ref{ass:inf_hor_lp} is automatically satisfied for all $k \geq k_0$ since the simulated ROM is controlled by the OCP \eqref{eq:rompc} and $\bar{\mathcal{Z}} \subseteq \mathcal{Z}$ and $\bar{\mathcal{U}} \subseteq \mathcal{U}$. Further, Assumption \ref{ass:inf_hor_lp} is easily verified in practice since the ROM is a simulated (non-physical) system. Assumption \ref{ass:epsilon0} is hard to verify since it would require knowledge of the full order state $x^f$. Nonetheless, as will be seen in the following sections, the value $\eta_{\underline{k}}$ can be chosen in an extremely conservative manner without having a major impact on the error bounds $\Delta_z$ and $\Delta_u$. This is accomplished by choosing the horizon parameter $\tau$ to be sufficiently large, which will be shown to significantly reduce the influence of $\epsilon_{\underline{k}}$ on $\epsilon_k$ for $k \geq k_0$. Additional discussion on the choice of $\tau$ is presented in Section \ref{subsec:eboundresults} and the choice of the matrix $G$ is discussed in Section \ref{subsec:computeMgGCrCw}.

\subsection{Computing Error Bounds}
The proposed methodology for computing the error bounds $\Delta_z$ and $\Delta_u$ used in \eqref{eq:tightconstraints} is now presented. First, note that the error $\epsilon_k$ is almost entirely defined by the most recent disturbances from $r$ and $\omega$ since the effects of older disturbances decay exponentially (when $A_\epsilon$ is Schur stable). This fact is exploited to yield a methodology that is scalable to high-dimensional problems while not being overly conservative. In particular, using \eqref{eq:error_response} the error $\epsilon_k$ is divided into two components $\epsilon_{k} = \epsilon_k^{(1)} + \epsilon_k^{(2)}$ where
\begin{equation*}
\begin{split}
\epsilon_k^{(1)} &= A_\epsilon^{k-\underline{k}} \epsilon_{\underline{k}} + \sum_{j=\underline{k}}^{k-\tau-1}A_\epsilon^{k-1-j} \Big(B_\epsilon r_j + G_\epsilon \omega_j \Big), \\
\epsilon_k^{(2)} &= \sum_{j=k - \tau}^{k-1}A_\epsilon^{k-1-j} \Big(B_\epsilon r_j + G_\epsilon \omega_j \Big).
\end{split}
\end{equation*}
The term $\epsilon_k^{(2)}$ represents the contribution from the $\tau$ most recent inputs, and the term $\epsilon_k^{(1)}$ represents the contribution from everything prior. By choosing $\tau$ to be sufficiently large, the error term $\epsilon_k^{(1)}$ can be made negligible, and therefore can be bounded with more conservative techniques without making the total bound more conservative. The dominant error $\epsilon_k^{(2)}$ is then analyzed using a worst-case optimization formulation that yields a tight bound.

\subsubsection{Bounding \texorpdfstring{$\epsilon_k^{(1)}$}{(e1)}}
This term is bounded by considering the \textit{norm} of the error, $\lVert \epsilon_k^{(1)} \rVert$. 
Using the triangle inequality and definition of induced matrix norms an upper bound on the weighted norm of $\epsilon_k^{(1)}$ can be expressed as
\begin{equation*}
\begin{split}
\lVert \epsilon_k^{(1)} \rVert_G \leq& \lVert A_\epsilon^{k-\underline{k}}\rVert_G \lVert \epsilon_{\underline{k}}\rVert_G + \sum_{j=\underline{k}}^{k-\tau-1}\lVert A_\epsilon^{k-1-j} \rVert_G \lVert B_\epsilon r_j\rVert_G \\
& +\sum_{j=\underline{k}}^{k-\tau-1}\lVert A_\epsilon^{k-1-j}\rVert_G \lVert G_\epsilon \omega_j\rVert_G, \\
\end{split}
\end{equation*}
where $G$ is the same positive definite matrix used to define the weighted norm in Assumption \ref{ass:epsilon0}.

Now, consider bounds $C_r$ and $C_\omega$ such that $\lVert B_\epsilon r_k\rVert_G \leq C_r$ and $\lVert G_\epsilon \omega_k\rVert_G \leq C_\omega$ for all $k \geq \underline{k}$, and parameters $M \geq 1$ and $\gamma \in (0,1)$ such that $\lVert A_\epsilon^i \rVert_G \leq M \gamma^i$ for all $i \geq 0$.
Since $A_\epsilon$ is Schur stable, parameters $M$ and $\gamma$ that meet these conditions are guaranteed to exist and $C_r$ and $C_\omega$ can be computed based on Assumptions \ref{ass:compact} and \ref{ass:inf_hor_lp}. Methods for computing these parameters are discussed in Sections \ref{subsubsec:CrandCw} and \ref{subsubsec:Mg}. These definitions lead to the bound
\begin{equation*}
\begin{split}
\lVert \epsilon_k^{(1)} \rVert_G \leq M\gamma^{k-\underline{k}} \eta_{\underline{k}} + M(C_r + C_\omega)\sum_{j=\underline{k}}^{k-\tau-1} \gamma^{k-1-j}. \\
\end{split}
\end{equation*}
Since $\lvert \gamma \rvert < 1$, for all $k \geq k_0$ the partial series is bounded by
\begin{equation*}
\sum_{j=\underline{k}}^{k-\tau-1}\gamma^{k-1-j} = \gamma^\tau \sum_{j=0}^{k-\tau-1-\underline{k}}\gamma^j \leq \gamma^\tau \sum_{j=0}^{\infty}\gamma^j = \frac{\gamma^\tau}{1-\gamma}.
\end{equation*}
Additionally, since $k - \underline{k} \geq 2\tau$ for all $k \geq k_0$ it holds that $\gamma^{k-\underline{k}} \leq \gamma^{2\tau}$ for all $k \geq k_0$. Finally, the upper bound $\Delta^{(1)}$ on the weighted norm of $\epsilon_k^{(1)}$ can be expressed as
\begin{equation} \label{eq:inf_hor_lp_e1}
\begin{split}
\lVert \epsilon_k^{(1)} \rVert_G \leq \Delta^{(1)} \coloneqq M\gamma^{2\tau}\eta_{\underline{k}} +\frac{M\gamma^\tau(C_r + C_\omega) }{1-\gamma},\\
\end{split}
\end{equation}
and $\eta_{\underline{k}}$ is defined in Assumption \ref{ass:epsilon0}. 
% It is important to note that $\Delta^{(1)}$ is \textit{constant} for all $k \geq k_0$.

\subsubsection{Bounding \texorpdfstring{$\epsilon_k^{(2)}$}{(e2)}}
The dominant error $\epsilon_k^{(2)}$ is considered next, and it can be noted that of specific interest is the error corresponding to the terms $E_z \epsilon_k^{(2)}$ and $E_u \epsilon_k^{(2)}$. Consider a specific row of either $E_z$ or $E_u$, denoted by $\theta^T$. A bound $\Delta^{(2)}(\theta)$, such that $\theta^T\epsilon_k^{(2)} \leq \Delta^{(2)}(\theta)$ for all $k \geq k_0$, can be computed via a worst-case optimization formulation
\begin{equation} \label{eq:inf_hor_lp_e2}
\begin{split} 
\Delta^{(2)}(\theta)= \underset{\bar{x}, \bar{u}, \omega}{\text{max.}} \quad &\theta^T \sum_{j=0}^{\tau-1}A_\epsilon^{\tau-j-1}\big(B_\epsilon r_j + G_\epsilon \omega_j\big), \\
\text{subject to} \quad & \bar{x}_{i+1} = A\bar{x}_i + B\bar{u}_i,\\
&r_i \in \bar{\mathcal{X}} \times \mathcal{U}, \:\: r_i = [\bar{x}_i^T, \bar{u}_i^T]^T, \\
&H\bar{x}_i \in \mathcal{Z}, \quad i \in [-\tau, \dots, \tau-1], \\
&\omega_i \in \mathcal{W} \times \mathcal{V}, \quad i \in [0,\dots, \tau-1],
\end{split}
\end{equation}
where $\bar{\mathcal{X}} \coloneqq \{ \bar{x} \:|\: H_{\bar{x}} \bar{x} \leq b_{\bar{x}} \}$ is a compact polytopic set that is used to guarantee that the problem is bounded. A method for computing $\bar{\mathcal{X}}$ is defined in Section \ref{subsubsec:Xbar}.

\subsubsection{Defining \texorpdfstring{$\Delta_z$}{(Dz)} and \texorpdfstring{$\Delta_u$}{(Du)}}
The bounds on the terms $\epsilon_k^{(1)}$ and $\epsilon_k^{(2)}$ are now combined to give the final bounds $\Delta_z$ and $\Delta_u$. Again considering a row $\theta^T$ of either $E_z$ or $E_u$:
\begin{equation*}
\begin{split}
\theta^T \epsilon_k &= \theta^T(\epsilon^{(1)}_k + \epsilon^{(2)}_k)\\
&\leq \lVert \theta^T G^{-1/2} \rVert_2 \lVert \epsilon^{(1)}_k \rVert_G + \theta^T \epsilon^{(2)}_k.\\
\end{split}
\end{equation*}
Therefore, from the previously computed bounds the $i$-th element of vector $\Delta_z$ and the $j$-th element of vector $\Delta_u$ are defined as
\begin{equation} \label{eq:bounds}
\begin{split}
\Delta_{z,i} &\coloneqq \Delta^{(1)}(e_{z,i}) + \Delta^{(2)}(e_{z,i}), \\
\Delta_{u,j} &\coloneqq \Delta^{(1)}(e_{u,j}) + \Delta^{(2)}(e_{u,j}), \\
\end{split}
\end{equation}
where $i \in [1,\dots,n_z]$, $j \in [1,\dots,n_u]$, $e_{z,i}^T$ and $e_{u,j}^T$ are the rows of $E_z$ and $E_u$ respectively, and
\begin{equation}
\begin{split}
\Delta^{(1)}(\theta) = \lVert \theta^T G^{-1/2} \rVert_2 \Delta^{(1)}.
\end{split}
\end{equation}

\subsection{Computing \texorpdfstring{$M, \gamma,G,C_r,C_\omega,$ and $\bar{\mathcal{X}}$}{(M,gamma,Cr,Cw,Xbar)}} \label{subsec:computeMgGCrCw}
The computation of several additional quantities are required to define the error bounds, namely $M$, $\gamma$, $G$, $C_r$, and $C_\omega$ for $\Delta^{(1)}$ in \eqref{eq:inf_hor_lp_e1}, and $\bar{\mathcal{X}}$ for $\Delta^{(2)}$ in \eqref{eq:inf_hor_lp_e2}. Methods for defining these parameters are now presented.

\subsubsection{Computing \texorpdfstring{$\bar{\mathcal{X}}$}{Xbar}} \label{subsubsec:Xbar}
The set $\bar{\mathcal{X}} \coloneqq \{ \bar{x} \:|\: H_{\bar{x}} \bar{x} \leq b_{\bar{x}} \}$ is used to ensure the linear programs \eqref{eq:inf_hor_lp_e2} are bounded, and will also be used in computing $C_r$. For simplicity this set is chosen to be a hyper-rectangle where the rows are the standard basis vectors in $\mathbb{R}^n$ (i.e. $H_{\bar{x}} = [I, -I]^T$). Further, define the $l$-th element of the vector $b_{\bar{x}}$ by the solution to the linear program
\begin{equation} \label{eq:Xbar}
\begin{split} 
b_{\bar{x},l} = \underset{\bar{x}, \bar{u}}{\text{maximize}} \quad &h_{\bar{x},l}^T \bar{x}_0, \\
\text{subject to} \quad &\bar{x}_{i+1} = A\bar{x}_i + B\bar{u}_i, \\
&\bar{u}_i \in \mathcal{U}, \quad i \in [0,\dots,\bar{i}-1] \\
&H\bar{x}_i \in \mathcal{Z}, \quad i \in [0,\dots,\bar{i}]
\end{split}
\end{equation}
where $h_{\bar{x},l}^T$ is the $l$-th row of $H_{\bar{x}}$ and $\bar{i}\geq n-1$.
With this definition of $\bar{\mathcal{X}}$ the following property holds:
\begin{prop} \label{prop:Xbar}
Suppose Assumptions \ref{ass:compact}, \ref{ass:ctrlobsv}, and \ref{ass:inf_hor_lp} hold. Then, the set $\bar{\mathcal{X}}$ is compact and $\bar{x}_k \in \bar{\mathcal{X}}$ for all $k \geq \underline{k}$.
\end{prop}

\subsubsection{Computing \texorpdfstring{$(C_r, C_\omega)$}{(Cr,Cw)}} \label{subsubsec:CrandCw}
The parameters $C_r$ and $C_\omega$ are used to bound the quantities $\lVert B_\epsilon r_k \rVert_G$ and $\lVert G_\epsilon \omega_k \rVert_G$, respectively. These bounds are computed by solving
\begin{equation} \label{eq:CrandCw}
\begin{split}
C_r &= \underset{\bar{x}\in \bar{\mathcal{X}}, \bar{u} \in \mathcal{U}}{\text{maximize}} \quad \lVert B_\epsilon [\bar{x}^T, \bar{u}^T]^T \rVert_G, \\
C_\omega &= \underset{w\in \mathcal{W}, v \in \mathcal{V}}{\text{maximize}} \quad \lVert G_\epsilon [w^T, v^T]^T \rVert_G,
\end{split}
\end{equation}
where $G$ is the same weighting matrix from Assumption \ref{ass:epsilon0}.
The optimization problems \eqref{eq:CrandCw} can be solved by vertex enumeration (since the constraints are defined by convex polytopes) or upper bounded by convex relaxations \cite{MangasarianShiau1986}.

\begin{prop} \label{prop:CrandCw}
Suppose Assumptions \ref{ass:compact}, \ref{ass:ctrlobsv}, and \ref{ass:inf_hor_lp} hold. Then, $\lVert B_\epsilon r_k \rVert_G \leq C_r$ and $\lVert G_\epsilon \omega_k \rVert_G \leq C_\omega$ for all $k \geq \underline{k}$, and $C_r$ and $C_\omega$ are finite.
\end{prop}

\subsubsection{Computing \texorpdfstring{$(M, \gamma,G)$}{(M,gamma,G)}} \label{subsubsec:Mg}
The parameters $M$ and $\gamma$ used in \eqref{eq:inf_hor_lp_e1} are required to satisfy $M \geq 1$ and $\gamma \in (0,1)$ and $\lVert A_\epsilon^i \rVert_G \leq M \gamma^i$ for all $i\geq 0$. Two potential approaches for computing these parameters are discussed here. 
First, if $G$ is chosen such that $\lVert A_\epsilon \rVert_G < 1$ then $\gamma = \lVert A_\epsilon \rVert_G$ and $M= 1$ can be used since $\lVert A_\epsilon^i \rVert_G \leq \lVert A_\epsilon \rVert_G^i$. A value of $G$ for this approach can be computed by solving the discrete Lyapunov equation
\begin{equation} \label{eq:lyap}
A_\epsilon^T G A_\epsilon - \eta^2 G + I = 0,
\end{equation}
where $\eta \in \mathbb{R}$ is some value such that $\eta \in ( \max_j \lvert \lambda_j(A_\epsilon)\rvert, 1)$ where $\lambda_j(A_\epsilon)$ is the $j$-th eigenvalue of $A_\epsilon$. This approach will guarantee that $\gamma \leq \eta$.
Second, if $A_\epsilon$ is diagonalizable with eigenvalue decomposition $A_\epsilon = T D T^{-1}$ then $\gamma = \max_j \lvert d_j\rvert$ and $M = \lVert G^{1/2} T \rVert_2 \lVert T^{-1}G^{-1/2} \rVert_2$ can be used, where $d_j$ denotes the $j$-th diagonal element of $D$. In this case $\lvert \gamma \rvert < 1$ is guaranteed by the Schur stability of $A_\epsilon$. In this case $G$ can be chosen to be any positive definite matrix, including simply $G=I$. However $G$ can also be more carefully selected to try to minimize the constant $M$ or more generally to decrease $\Delta^{(1)}$. This can be accomplished via an optimization based approach \cite[Section~V]{LorenzettiPavone2020}.

\subsection{Results and Practical Considerations} \label{subsec:eboundresults}
The procedure for computing the error bounds $\Delta_z$ and $\Delta_u$ that are used to tighten the constraints \eqref{eq:tightconstraints} is summarized in Algorithm \ref{alg:ebounds}.
\begin{algorithm}
\caption{Error Bound Analysis}\label{alg:ebounds}
\begin{algorithmic}[1]
\Procedure{ErrorBounds}{$\eta_{\underline{k}}$, $\tau$, $\bar{i}$, $\dots$}
\State Compute $\bar{\mathcal{X}}$ (Section \ref{subsubsec:Xbar})
\State Compute $(C_r, C_\omega)$ (Section \ref{subsubsec:CrandCw})
\State Compute $(M, \gamma,G)$ (Section \ref{subsubsec:Mg})
\State Compute $\Delta^{(1)}$ \eqref{eq:inf_hor_lp_e1}
\State Compute $\Delta^{(2)}(\cdot)$ \eqref{eq:inf_hor_lp_e2}
\State Compute $\Delta_z$, $\Delta_u$ \eqref{eq:bounds} 
\State \Return $\Delta_z$, $\Delta_u$
\EndProcedure
\end{algorithmic}
\end{algorithm}
With the definition of these error bounds, Lemma \ref{lem:robustconstraint} can be extended to the following result:
\begin{thm}[Robust Constraint Satisfaction] \label{thm:robustconstraint}
Suppose Assumptions \ref{ass:compact}-\ref{ass:inf_hor_lp} hold and that at time $k_0$ the OCP \eqref{eq:rompc} is feasible. Then, under the proposed controller the full order system will robustly satisfy the constraints \eqref{eq:constraints} for all $k \geq k_0$.
\end{thm}

Results from applying this error bounding procedure to the example problems discussed in Section \ref{sec:experiments} are now presented. In particular the following values are presented in Table \ref{tab:ebounds} (for the disturbance free case):
\begin{enumerate}
\item $r = \max\{r_z,  r_u\}$ where $$r_{(z,u)} = \max_i 100 \times \frac{\Delta^{(1)}(e_{(z,u),i})}{b_{(z,u),i}},$$
\item $t_{(z,u),\text{max}} = \max_i \tilde{\Delta}_{(z,u),i}$ and $t_{(z,u),\text{min}} = \min_i \tilde{\Delta}_{(z,u),i}$ where $$\tilde{\Delta}_{(z,u),i} = 100 \times \frac{\Delta_{(z,u),i}}{b_{(z,u),i}}.$$
\end{enumerate}
The value $r$ defines the maximum percentage (over all constraints on $z$ and $u$) of constraint tightening induced by $\Delta^{(1)}$ with respect to the original constraint values $b_z$ and $b_u$ as defined in \eqref{eq:constraints}. In other words, a value of $r=1$ means that over all constraints, the $\Delta^{(1)}$ error would cause the constraint to be tightened by \textit{at most} $1\%$.
The values $\tilde{\Delta}_{(z,u),i}$ define the percentage of \textit{total} constraint tightening for each constraint, with respect to the constraint value. Therefore $t_{(z,u),\text{max}}$ is the maximum percentage of constraint tightening for either the constraints on $z$ or $u$, and similarly $t_{(z,u),\text{min}}$ is the minimum. For example, a value of $t_{z,\text{max}} = 1$ means that when defining the tightened constraint set $\bar{\mathcal{Z}}$, the most a constraint is tightened by is $1\%$ of its original value.

\begin{table}[htbp]
\caption{Error Bound Results}
\centering
\setlength\tabcolsep{5pt}
\renewcommand{\arraystretch}{1.2}
\begin{tabular}{|c|c|c|c|c|c|c|}
\cline{1-7}
\multicolumn{1}{|r|}{Model} & $\tau$ &  $r$ & $t_{z,\text{max}}$ & $t_{z,\text{min}}$ & $t_{u,\text{max}}$ & $t_{u,\text{min}}$  \\ \hline
\multicolumn{1}{|r|}{Small Synthetic} & 200 & 3.7e-6 & 13.3 & 3.3 & 0.5 & 0.5  \\ \hline
\multicolumn{1}{|r|}{Large Synthetic} & 500 & 2.6e-3 & 4.1 & 4.1 & 6.7 & 6.7  \\ \hline
\multicolumn{1}{|r|}{Dist. Column} & 2000 & 3.7e-4 & 1.5 & 0.1 & 3.9 & 0.5   \\ \hline
\multicolumn{1}{|r|}{Tubular Reactor} & 1200 & 0.6 & 24.4 & 0.9 & 0.6 & 2.3e-3 \\ \hline
\multicolumn{1}{|r|}{Heatflow} & 1500 & 0.4 & 41.7 & 20.6 & 1.5 & 0.5 \\ \hline
\multicolumn{1}{|r|}{Sup. Diffuser} & 2000 & 3.9e-9 & 0.2e-4 & 0.1e-4 & 0.4e-6 & 0.1e-6 \\ \hline
\multicolumn{7}{p{240pt}}{Results from computing the error bounds for examples described in Section \ref{sec:experiments} with no external disturbances (model reduction error only). The parameter $\tau$ is the time horizon and in each case $\eta_{\underline{k}} = 10^{10}$ and $\bar{i} = \tau$. The reported quantity $r$ is the maximum percentage of constraint tightening resulting from just the $\Delta^{(1)}$ term in \eqref{eq:bounds}. The quantities $t_{(z,u),(\text{max},\text{min})}$ are the max/min percentage of $z$ and $u$ constraint tightening resulting from the entire bounds $\Delta_z$ and $\Delta_u$ defined by \eqref{eq:bounds}. }\\
\end{tabular}
\label{tab:ebounds}
\end{table}

\begin{figure}[!t]
\centerline{\includegraphics[width=0.85\columnwidth]{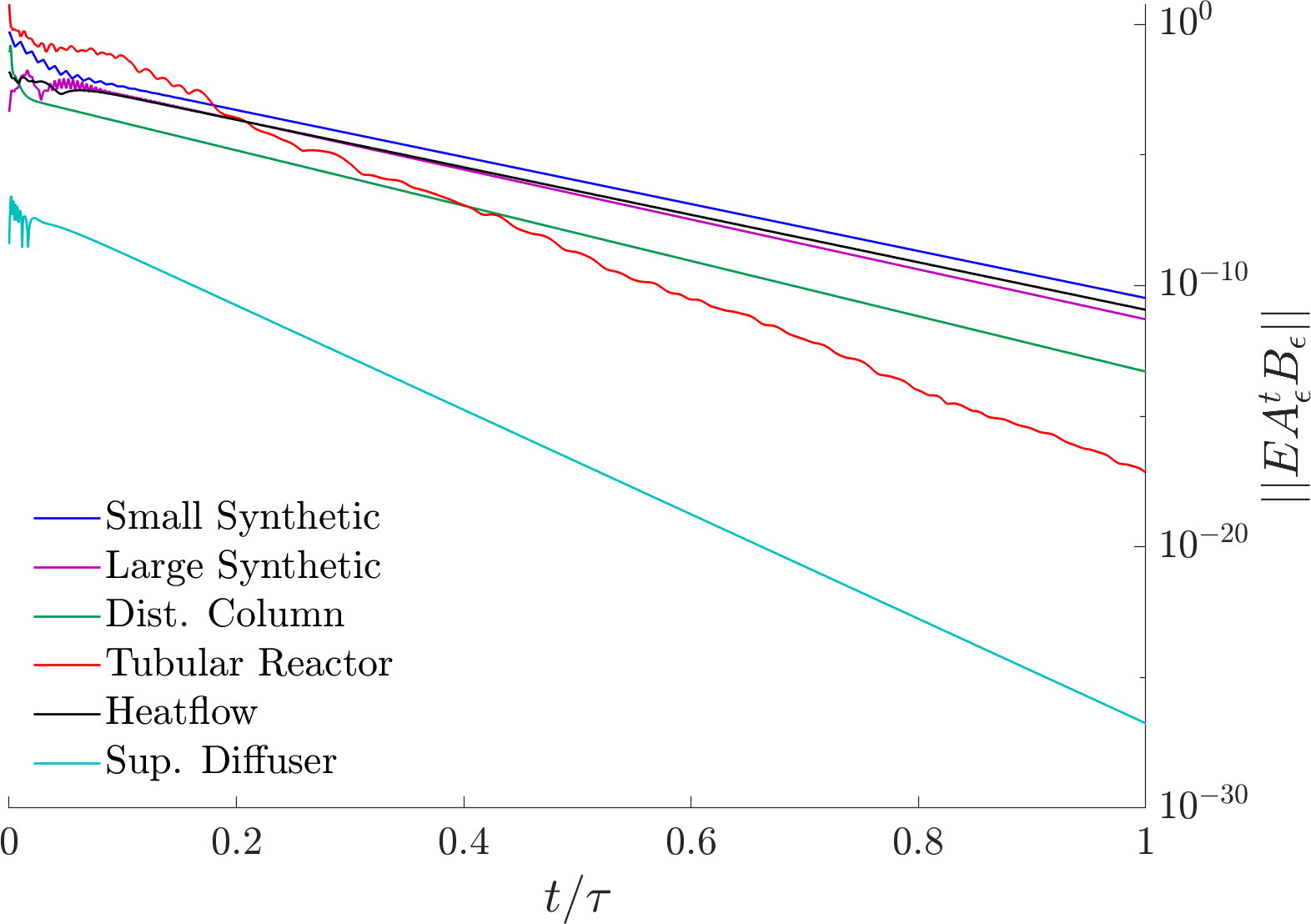}}
\caption{Plot showing the matrix norm $\lVert E A_\epsilon^t B_\epsilon \rVert$, where $E = [E_z^T, E_u^T]^T$, for different values of $t \in [0,\dots,\tau]$. For each example the value of $\tau$ is the same as presented in Table \ref{tab:ebounds}.}
\label{fig:emat_norm_decay}
% \vspace{-10pt}
\end{figure}

The values of $r$ are presented to demonstrate how the error bound component from $\Delta^{(1)}$ ends up being only a small fraction of the total error bound. This occurs because a large value of $\tau$ leads to the vast majority of the error being captured by the bound $\Delta^{(2)}$. As mentioned before, this reduces conservatism since the norm bound approach used to define $\Delta^{(1)}$ is naturally more conservative. Another simple experiment to demonstrate this phenomenon is to study the change of the matrix norm $\lVert E A_\epsilon^t B_\epsilon \rVert$, where $E = [E_z^T, E_u^T]^T$, as $t$ increases, which is shown in Figure \ref{fig:emat_norm_decay} for $t \in [0,\dots,\tau]$. As can be seen, for the values of $\tau$ in Table \ref{tab:ebounds} the matrix norms decrease exponentially, and for $t = \tau$ are all less than $10^{-10}$. Therefore larger values of $\tau$ would have negligible effect on the error bounds. This analysis could also be used to provide a more automated way to choose the parameter $\tau$, for example by choosing $\tau$ such that $\lVert E A_\epsilon^t B_\epsilon \rVert \leq \epsilon_{\text{threshold}}$.  

\subsubsection{Computational Efficiency}
For extremely high-dimensional problems, computing the $\Delta^{(1)}$ term may be challenging or impossible (in particular computing the bounds on the norm of the matrix powers $\lVert A_\epsilon^i \rVert_G \leq M \gamma^i$). However, as is noted in Remark \ref{rem:ebounds}, it is significantly more important in practice to be able to compute the bounds $\Delta^{(2)}$. In fact, computing the bounds $\Delta^{(2)}$ remains computationally tractable since the complexity of the linear programs \eqref{eq:inf_hor_lp_e2} only scale with the size of the reduced order model $n$, the number of control inputs $m$, the dimensions of the disturbances $m_w$ and $p$, and the time horizon parameter $\tau$. In many practical problems these parameters will be small enough that \eqref{eq:inf_hor_lp_e2} can be solved efficiently with standard commercial solvers.
Additionally, the number of linear programs that need to be solved only scales with the number of constraints that define $\mathcal{Z}$ and $\mathcal{U}$, which is typically not large.

One potentially challenging component to solving the linear programs \eqref{eq:inf_hor_lp_e2} is computing the matrices $E A_\epsilon^i B_\epsilon$ and $E A_\epsilon^i G_\epsilon$ for $i = 0,\dots,\tau$ (that define the objective functions) since the high-dimensional matrix $A_\epsilon$ is involved. However, the dimension of the resulting matrix products do not scale with $n^f$, and can be computed efficiently in a recursive fashion as long as the matrix-vector product operation can be efficiently performed (e.g. if $A_\epsilon$ is sparse). In fact this is equivalent to simulating the dynamics from a particular initial condition.

\begin{rem}[Practical Considerations for Extremely High-Dimensional Problems] \label{rem:ebounds}
Note that the preceding discussion and analysis (Figure \ref{fig:emat_norm_decay}) suggests that choosing $\tau$ to be large makes $\Delta^{(1)}$ negligible. Therefore, from a \textit{practical} standpoint the proposed approach can be scaled to extremely high-dimensional problems. In particular, the computation of $\Delta^{(1)}$ could be foregone by choosing $\tau$ to be large enough such that $\lVert E A_\epsilon^\tau B_\epsilon \rVert$ is small. In fact, this is the approach taken for the Aircraft example in Section \ref{subsec:aircraft}, which has a model dimension of $n^f = 998,936$.
\end{rem}

\section{Constant Setpoint Tracking} \label{sec:setpointtracking}
The ROMPC scheme presented in Section \ref{sec:rompc} can also be extended to setpoint tracking problems, where a subset of the performance variables $z$ are controlled to track a constant reference signal.
However special care must be taken to avoid steady-state tracking errors that are induced by model reduction error, which we discuss in this section.

Consider the tracking variables $z^r := Tz$ where the matrix $T \in \mathbb{R}^{t \times o}$ is defined by taking the $t$ rows of the identity matrix corresponding to the indices of the desired performance variables. The goal is to make $z^r$ converge to a constant reference signal $r$:
\begin{equation}
\lim_{k \rightarrow \infty }z^r_k = r. 
\end{equation}

First, the FOM equilibrium corresponding to the setpoint $r$ can be computed by solving the linear system 
\begin{equation} \label{eq:fomss}
S^f \begin{bmatrix}
x^f_{\infty} \\ u_{\infty}
\end{bmatrix} = 
\begin{bmatrix}
0 \\ r
\end{bmatrix}, \quad S^f=\begin{bmatrix}
A^f - I & B^f \\ TH^f & 0
\end{bmatrix}.
\end{equation}
To ensure that \eqref{eq:fomss} has a unique solution, it is assumed that the matrix $S^f$ is square (i.e., number of tracking variables $t$ is equal to the number of control inputs $m$), and full rank.

\subsection{Setpoint Tracking ROMPC}
Modifying the ROMPC scheme defined in Section \ref{sec:rompc} for setpoint tracking only requires slight changes to the OCP \eqref{eq:rompc}, namely in the cost function and terminal constraint set. First, the cost function in \eqref{eq:rompc} should be changed to
\begin{equation}
\lVert \delta \bar{x}_{k+N|k} \rVert^2_P + \sum_{j=k}^{k+N-1}\lVert \delta \bar{x}_{j|k}\rVert^2_Q + \lVert \delta \bar{u}_{j|k}\rVert^2_R,
\end{equation}
where $\delta \bar{x} = \bar{x} - \bar{x}_\infty$ and $\delta \bar{u} = \bar{u} - \bar{u}_\infty$ and $(\bar{x}_\infty, \bar{u}_\infty)$ are the desired steady state targets for the simulated ROM. Second, the terminal set $\bar{\mathcal{X}}_f$ should be computed using the same techniques discussed in Section \ref{subsec:roocp}, but should be designed around the target point $(\bar{x}_\infty, \bar{u}_\infty)$ rather than the origin (see \cite[Section~IV]{LorenzettiLandryEtAl2019}). The combination of these changes is sufficient to guarantee that the simulated ROM controlled by the reduced order OCP exponentially converges to the target point $(\bar{x}_\infty, \bar{u}_\infty)$ (rather than the origin).

Due to model reduction errors, the target point $(\bar{x}_\infty, \bar{u}_\infty)$ needs to be carefully chosen to ensure setpoint tracking for the controlled FOM. For a setpoint $r$ and associated solution $(x^f_\infty, u_\infty)$ to \eqref{eq:fomss}, the reduced order estimator steady state value (ignoring disturbances) is given by
\begin{equation} \label{eq:estss}
\hat{x}_{\infty} = D(Bu_{\infty} + LC^fx^f_{\infty}),
\end{equation}
where $D = (I-(A-LC))^{-1}$.
Then, by requiring the controller (\ref{eq:controller}) to also be at steady state, the target point $(\bar{x}_\infty, \bar{u}_\infty)$ that enables setpoint tracking is the solution to
\begin{equation} \label{eq:romss}
S \begin{bmatrix}
\bar{x}_{\infty} \\ \bar{u}_{\infty}
\end{bmatrix} = 
\begin{bmatrix}
0 \\ K\hat{x}_{\infty} - u_{\infty}
\end{bmatrix}, \quad S = \begin{bmatrix}
A-I & B\\ K & -I
\end{bmatrix},
\end{equation}
where it is assumed that the square matrix $S$ is full rank.
Of course both $(x^f_\infty, u_\infty)$ and $(\bar{x}_\infty, \bar{u}_\infty)$ must also satisfy the constraints $z_\infty \in \mathcal{Z}$, $u_\infty \in \mathcal{U}$, $\bar{z}_\infty \in \bar{\mathcal{Z}}$, $\bar{u}_\infty \in \bar{\mathcal{U}}$.

\subsection{Setpoint Tracking Closed-Loop Performance}
The same closed-loop performance as in Section \ref{sec:closedloopperf} is achieved, except that the system asymptotically converges to $x^f_\infty$ rather than the origin. In particular Lemma \ref{lem:robustconstraint} still holds, which guarantees robust constraint satisfaction.

\begin{thm}[Robust Setpoint Tracking] \label{thm:robusttracking}
Suppose that $A_\epsilon$ is Schur stable, and that the OCP is feasible at time $k_0$ and drives the simulated ROM to the target point $(\bar{x}_\infty, \bar{u}_\infty)$ exponentially fast. Additionally, suppose the target point is defined for some setpoint $r$ such that $(x^f_\infty, u_\infty)$, $\hat{x}_\infty$, and $(\bar{x}_\infty, \bar{u}_\infty)$ are the unique solutions to the equations \eqref{eq:fomss}, \eqref{eq:estss}, \eqref{eq:romss}.
Then, the closed-loop system robustly and asymptotically converges to a compact set containing $x^f_\infty$ and the tracking variables $z^r$ converge to a compact set containing $r$.
\end{thm}

\begin{crl}[Setpoint Tracking] \label{crl:tracking}
Let the conditions from Theorem \ref{thm:robusttracking} hold.
Then, if $\mathcal{W} =\{0\}$ and $\mathcal{V} = \{0\}$ the tracking variables $z^r$ converge asymptotically to the setpoint $r$.
\end{crl}

\section{Continuous-time ROMPC} \label{sec:continuoustime}
In some practical settings a continuous time formulation of the problem might be preferable, for example to avoid discretizing a high-dimensional continuous-time FOM in time. For the benefit of the practitioner, this section includes a discussion on how the proposed ROMPC scheme can be applied to continuous-time systems with only a few modifications.

First, the FOM \eqref{eq:fom} and ROM \eqref{eq:rom} would be expressed as ordinary differential equations instead of difference equations, and the cost functions \eqref{eq:fomcost} and \eqref{eq:romcost} would be expressed as integrals.
Second, the continuous-time versions of the control law \eqref{eq:controller} and state estimator \eqref{eq:estimator} are defined by
\begin{align}
    u(t) &= \bar{u}(t) + K(\hat{x}(t) - \bar{x}(t)), \label{eq:controller:ct}\\
    \dot{\hat{x}}(t) &= A \hat{x}(t) + Bu(t) + L(y(t) - C\hat{x}(t)). \label{eq:estimator:ct}
\end{align}
The continuous-time simulated ROM is given by
\begin{equation} \label{eq:rom:ct}
\begin{split}
\dot{\bar{x}}(t) &= A \bar{x}(t) + B \bar{u}(t), \\
\bar{z}(t) &= H \bar{x}(t),
\end{split}
\end{equation}
and is controlled by the reduced order optimal control problem
\begin{equation} \label{eq:rompc:ct}
\begin{split}
(\bar{\mathbf{x}}^*_k, \bar{\mathbf{u}}^*_k) = \underset{\bar{\mathbf{x}}_k, \bar{\mathbf{u}}_k}{\text{argmin.}} \:\:& \lVert \bar{x}_{k+N|k} \rVert^2_P + \sum_{j=k}^{k+N-1}\lVert \bar{x}_{j|k}\rVert^2_Q + \lVert \bar{u}_{j|k}\rVert^2_R, \\
\text{subject to} \:\: & \bar{x}_{i+1|k} = A_d\bar{x}_{i|k} + B_d\bar{u}_{i|k},  \\
& H\bar{x}_{i|k} \in \bar{\mathcal{Z}}, \quad \bar{u}_{i|k} \in \bar{\mathcal{U}}, \\
& \bar{x}_{k+N|k} \in \bar{\mathcal{X}}_f,  \quad \bar{x}_{k|k} = \bar{x}(t),
\end{split}
\raisetag{3\normalbaselineskip}
\end{equation}
which is essentially the same as \eqref{eq:rompc}, but where the initial condition uses the continuous-time state $\bar{x}(t)$ and the dynamics model is a discretized version of the ROM \eqref{eq:rom:ct}. In particular, a zero-order hold equivalent ROM with $A_d \coloneqq e^{A\Delta t}$ and $B_d \coloneqq \int_0^{\Delta t} e^{A \tau}B d\tau $ is used, where $\Delta t$ is the discretization time. With this choice, the continuous-time simulated ROM will match the solution to the OCP at the sampled time steps under a zero-order hold control. In other words, suppose the OCP was solved with $\bar{x}_{k|k} = \bar{x}(t)$ to give the optimal discrete-time trajectory sequence $(\{\bar{x}^*_i\}_{i=k}^{k+N}, \{\bar{u}^*_i\}_{i=k}^{k+N-1})$. Then, if the control sequence is applied as a zero-order hold to the simulated ROM \eqref{eq:rom:ct} the trajectories would be equivalent at the sampled times: $\bar{x}(t + i\Delta t) = \bar{x}^*_{k+i}$ for $i = 0,\dots,N$. Note that the reduced order OCP is only solved every $\Delta t$ seconds.

The overall method is summarized in Algorithm \ref{alg:online:ct}, where it is assumed that the OCP is feasible at $t = t_0$ and where the control rate is given by $\delta t$ (which is generally faster than $\Delta t$).
\begin{algorithm}
\caption{ROMPC Control (Online, Continuous-Time)}\label{alg:online:ct}
\begin{algorithmic}[1]
\Procedure{ROMPC}{$\delta t$}
\State $t \leftarrow 0$, $\bar{x}(t) \leftarrow 0$, $\hat{x}(t) \leftarrow 0$
\Loop
\If{$t \geq t_0$}
    \If{$(t-t_0)\mod{\Delta t} = 0$}
    \State $\bar{u} \leftarrow$ solveOCP($\bar{x}(t)$)
    \EndIf
    \State $u(t) \leftarrow \bar{u} + K(\hat{x}(t) - \bar{x}(t))$
\Else
    \State $u(t) \leftarrow$ startupControl()
    \State $ \bar{u}(t) \leftarrow u(t) - K(\hat{x}(t) - \bar{x}(t))$
\EndIf
\State $y(t) \leftarrow$ getMeasurement()
\State applyControl($u(t)$)
\State $\bar{x}(t + \delta t) \leftarrow$ integrate \eqref{eq:rom:ct}
\State $\hat{x}(t + \delta t) \leftarrow$ integrate \eqref{eq:estimator:ct}
\State $t \leftarrow t + \delta t$
\EndLoop
\EndProcedure
\end{algorithmic}
\end{algorithm}

\subsection{Controller Synthesis and Error Bounds}
The reduced order Riccati method proposed in Section \ref{sec:controllersynth} can also be applied, but with the solution to the continuous algebraic Riccati equations
\begin{equation} \label{eq:h2riccati:ct}
\begin{split}
A^TX + XA - XBR^{-1}B^TX + Q_z &= 0, \\
AY + YA^T - YC^TCY + Q_w &= 0, \\
\end{split}
\end{equation}
and choosing $K = -R^{-1}B^TX$ and $L = YC^T$.

The continuous-time error dynamics can be written as
\begin{equation*} \label{eq:error_response:ct}
\epsilon(t) = e^{A_\epsilon (t-\underline{t}}\epsilon(\underline{t}) + \int_{\underline{t}}^t e^{A_\epsilon (t-s)}(B_\epsilon r(s) + G_\epsilon\omega(s)) ds,
\end{equation*}
where $\underline{t} = t_0 - 2\tau$.
Analogously to Assumptions \ref{ass:inf_hor_lp} and \ref{ass:epsilon0} it is assumed that $\bar{u}(t) \in \mathcal{U}$ and $\bar{z}(t) \in \mathcal{Z}$ for all $t \geq \underline{t}$ and that $\lVert \epsilon(\underline{t}) \rVert_G \leq \eta_{\underline{t}}$. The error is again divided into two components $\epsilon^{(1)}(t)$ and $\epsilon^{(2)}(t)$ with each term bounded separately. The bound on $\epsilon^{(1)}(t)$ is defined as
\begin{equation} \label{eq:inf_hor_lp_e1:ct}
\begin{split}
\lVert \epsilon^{(1)}(t) \rVert_G \leq \Delta^{(1)} \coloneqq \beta e^{\alpha(2\tau)}\eta_{\underline{t}} +\frac{\beta e^{\alpha \tau}(C_r + C_\omega) }{-\alpha},\\
\end{split}
\end{equation}
where $\beta$ and $\alpha$ are defined such that $\lVert e^{A_\epsilon t} \rVert_G \leq \beta e^{\alpha t}$ for all $t \geq 0$ and the values $C_r$ and $C_\omega$ are bounds such that $\lVert B_\epsilon r(t) \rVert_G \leq C_r$ and $\lVert G_\epsilon \omega(t) \rVert_G \leq C_\omega$. The values $C_r$ and $C_\omega$ can be computed via \eqref{eq:CrandCw} with $\bar{\mathcal{X}}$ defined as in Section \ref{subsubsec:Xbar} except with the discrete-time $A_d$ and $B_d$.

The second error term is given by
\begin{equation*}
\begin{split}
\epsilon^{(2)}(t) &= \int_{t-\tau}^t e^{A_\epsilon (t-s)}(B_\epsilon r(s) + G_\epsilon\omega(s)) ds. \\
\end{split}
\end{equation*}
which can be approximated via a quadrature scheme
\begin{equation*}
\begin{split}
\epsilon^{(2)}(t) &\approx \sum_{j=0}^{N_s} w_j e^{A_\epsilon (t-s_j)}(B_\epsilon r(s_j) + G_\epsilon\omega(s_j)), \\
\end{split}
\end{equation*}
where $N_s$ evenly spaced grid points are used such that $s_0 = t-\tau$, $s_1 = s_0 + \Delta t$ and so on.
Therefore the bounds $\Delta^{(2)}(\theta)$ can be computed by solving the linear programs
\begin{equation} \label{eq:inf_hor_lp_e2:ct}
\begin{split} 
\Delta^{(2)}(\theta)= \underset{\bar{x}, \bar{u}, \omega}{\text{max.}} \quad &\theta^T \sum_{j=0}^{N_s}w_j e^{A_\epsilon(\tau-j\Delta t)}\big(B_\epsilon r_j + G_\epsilon \omega_j\big), \\
\text{subject to} \quad & \bar{x}_{i+1} = A_d\bar{x}_i + B_d\bar{u}_i,\\
&r_i \in \bar{\mathcal{X}} \times \mathcal{U}, \:\: r_i = [\bar{x}_i^T, \bar{u}_i^T]^T, \\
&H\bar{x}_i \in \mathcal{Z}, \quad i \in [-N_s, \dots, N_s], \\
&\omega_i \in \mathcal{W} \times \mathcal{V}, \quad i \in [0,\dots, N_s].
\end{split}
\raisetag{3\normalbaselineskip}
\end{equation}
Finally, the combined bounds $\Delta_z$ and $\Delta_u$ for constraint tightening are computed by \eqref{eq:bounds}.

Note that in this case the guarantees on constraint satisfaction are made without accounting for the numerical approximations error (e.g. from approximation of the integral when considering $\epsilon^{(2)}(t)$). Accounting for this error to derive exact guarantees is left for future work. However, in practice this analysis is still useful for approximately quantifying the worst-case error. The computational efficiency is consistent with the discrete-time case. In particular the matrices that define the cost functions in \eqref{eq:inf_hor_lp_e2:ct}, namely $Ee^{A_\epsilon s}B_\epsilon$ and $Ee^{A_\epsilon s}G_\epsilon$ for $s = 0,\Delta t, 2\Delta t, \dots, \tau$, can be efficiently computed. One approach would be to first compute the matrix exponential $e^{A_\epsilon\Delta t}$ and then to recursively compute the matrix products, but for extremely large problems another approach would be to use some integration scheme (e.g. implicit Euler scheme) to simulate the dynamics
\begin{equation*}
\dot{\epsilon} = A_\epsilon \epsilon, \quad \epsilon(0) = b_\epsilon, g_\epsilon,
\end{equation*}
where $b_\epsilon$ and $g_\epsilon$ are the columns of $B_\epsilon$ and $G_\epsilon$.

\section{Simulation Results} \label{sec:experiments}
In this section a brief description of the models used to generate the results in Tables \ref{tab:ctrlsynth}, \ref{tab:ctrlsynthtimes}, and \ref{tab:ebounds} is provided. Complete examples for each model, including definitions of the FOM, ROM, constraints, and synthesis of a ROMPC controller are provided online\footnote{{https://github.com/StanfordASL/rompc}}. In-depth discussion and simulation results are then presented for two high-dimensional problems.

\subsection{Small and Large Synthetic Examples}
The small synthetic discrete-time system is a slightly modified version of the model defined in \cite{HovlandLovaasEtAl2008} with $n^f = 6$, $m=1$, $o=2$, and was also used in \cite{LorenzettiLandryEtAl2019}. The ROM is defined by balanced truncation and has dimension $n = 4$. The large synthetic model is also a discrete-time system, but with a dimension $n^f = 20$, $m=1$, $o=1$ and balanced truncation ROM of dimension $n=6$.

\subsection{Distillation Column}
This example is a linearized model of a distillation column used to separate a binary mixture into high purity products. The continuous-time model is defined in \cite{SkogestadPostlethwaite2005} (referred to as Column A). A discrete-time model with $n^f = 86$, $m=4$, $o=8$ is defined with a sampling time of one minute, where the original model inputs are appended to the state and their derivatives are controlled. This model has two integrating modes, and a balanced truncation ROM with $n=12$ is defined by performing a stable/unstable mode decomposition and only reducing the stable portion (i.e. the integrating modes are not reduced). 

\subsection{Tubular Reactor}
Tubular reactor models are used to study simplified chemical reaction control processes. This PDE model, defined in \cite{AgudeloEspinosaEtAl2007}, describes the change in reactant concentration and temperature along a one-dimensional flow. The model is discretized spatially with 300 nodes to produce a FOM with $n^f = 600$, $m=3$, $o=10$. In this example the continuous-time model is considered using the methods described in Section \ref{sec:continuoustime}. A balanced truncation model with dimension $n=30$ is used.

\subsection{Heatflow}
This example uses a PDE model describing the two-dimensional heat transfer over a flat plate that is spatially discretized. The model is a modified version of the HF2D9 model from \cite{Leibfritz2006} and is the same as the model used in \cite{LorenzettiPavone2020b}. The full order model is discretized in time, is unstable, and has dimension $n^f = 3481$, $m=5$, $o=4$. A stable/unstable decomposition is performed and the reduced order model with $n=21$ is defined by reducing the stable subsystem via balanced truncation.

\subsection{Supersonic Diffuser} \label{subsec:supdiff}
Supersonic diffusers, which slow down and increase pressure in compressible fluids moving at supersonic velocities, are commonly used in jet engines to decelerate incoming flow to subsonic speeds before entering the engine compressor. In nominal operating conditions a normal shock forms downstream of the diffuser throat section, yet if the flow is sufficiently disturbed the shock can move upstream of the throat and cause severe engine disruption.
This example studies a problem where an air bleed valve can be used to indirectly control the position of the normal shock. A CFD model with dimension $n^f = 11730$ is used, which is described in \cite{Lassaux2002, WillcoxLassaux2005, DavisHu2011}, and has also been used in the context of ROMPC in \cite{HovlandGravdahlEtAl2008}. The model has been time-discretized with a sampling time of $0.025$ seconds and a ROM is generated via balanced truncation with  $n=10$.

In this problem the nominal operating condition of the engine is at a Mach number of $M=2.2$ with a throat Mach number of $M_T = 1.36$ (the condition $M_T > 1$ ensures the normal shock occurs downstream of the throat). The nominal air bleed is $b = 0.01$, where $b$ is the ratio of bleed mass flow rate to engine airflow mass flow rate. The control inputs are the perturbations to the nominal air bleed, $\Delta b$, and the performance (and measured) output is the change in nominal throat Mach number, $\Delta M_T$. Constraints on these values are given by $-0.01 \leq \Delta b \leq 0.04$ and $-0.26 \leq \Delta M_T \leq 0.44$. A process disturbance is defined by perturbations in the incoming flow density that are assumed bounded $\lvert w \rvert \leq 1.225(10)^{-3}$ ($0.1\%$ of sea level density), and the measurement noise is assumed bounded by $\lvert v \rvert \leq 10^{-4}$.

The controller gains are computed using the reduced order Riccati method (Algorithm \ref{alg:controllersynth}) with $W_z = 1$ and $W_u = 0.5$. Results on the error bounds in the disturbance free case are given in Table \ref{tab:ebounds}, and with disturbances the constraints on $\Delta M_T$ are tightened by $0.6\%$ and $1.0\%$ for upper and lower bounds respectively, and the constraints on $\Delta b$ are tightened by $.01\%$ and $.05\%$. These error bounds are quite small, which is mainly due to the fact that the ROM is a very good approximation. The OCP was defined with a horizon $N = 20$.

\begin{figure}[ht]
\centerline{\includegraphics[width=0.75\columnwidth]{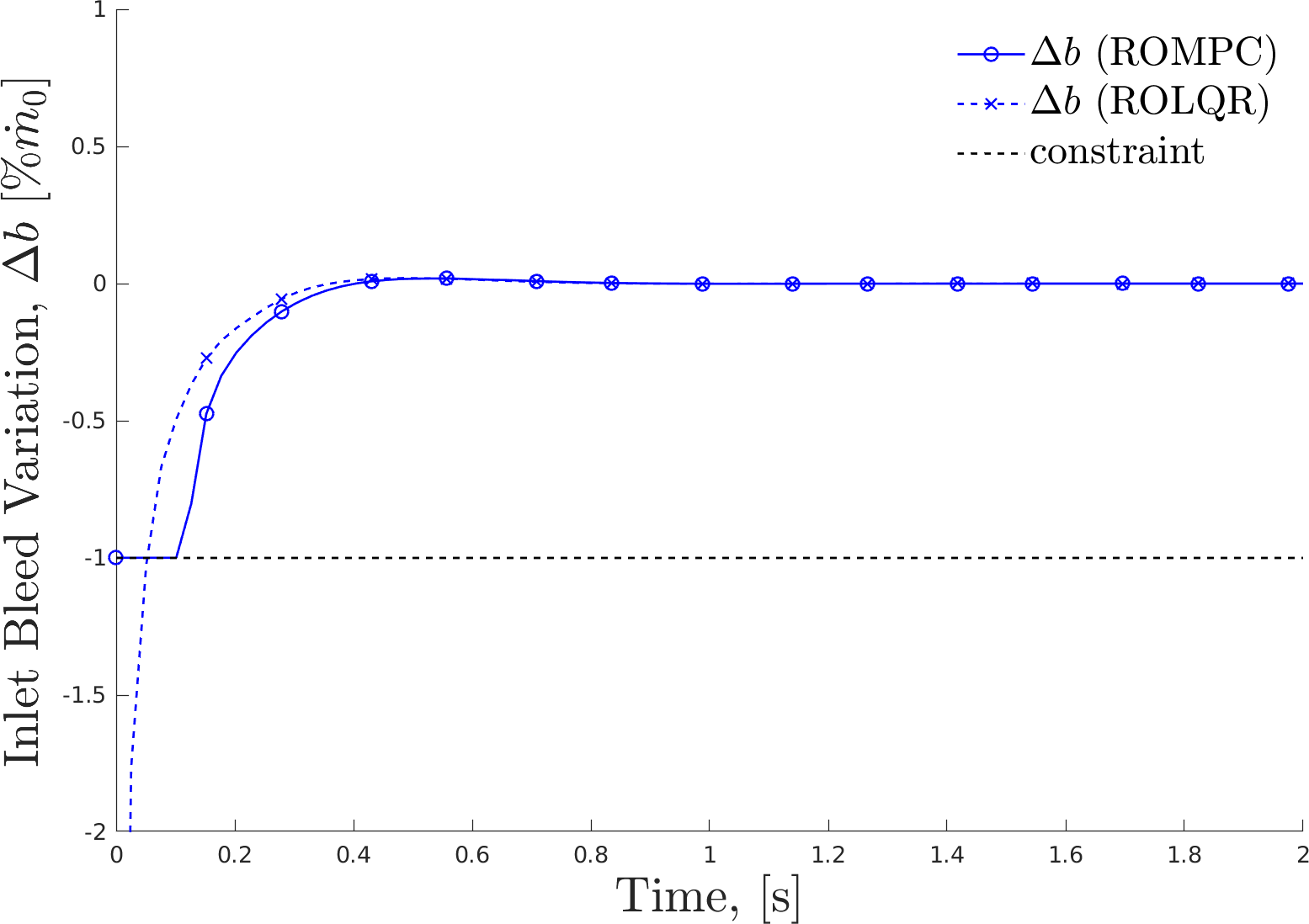}}
\caption{Simulation results for the supersonic diffuser example in Section \ref{subsec:supdiff}. This plot shows the control input for two simulations, one with the ROMPC scheme and another with a reduced order LQR controller.}
\label{fig:u_sup_diffuser}
\end{figure}

\begin{figure}[ht]
\centerline{\includegraphics[width=0.75\columnwidth]{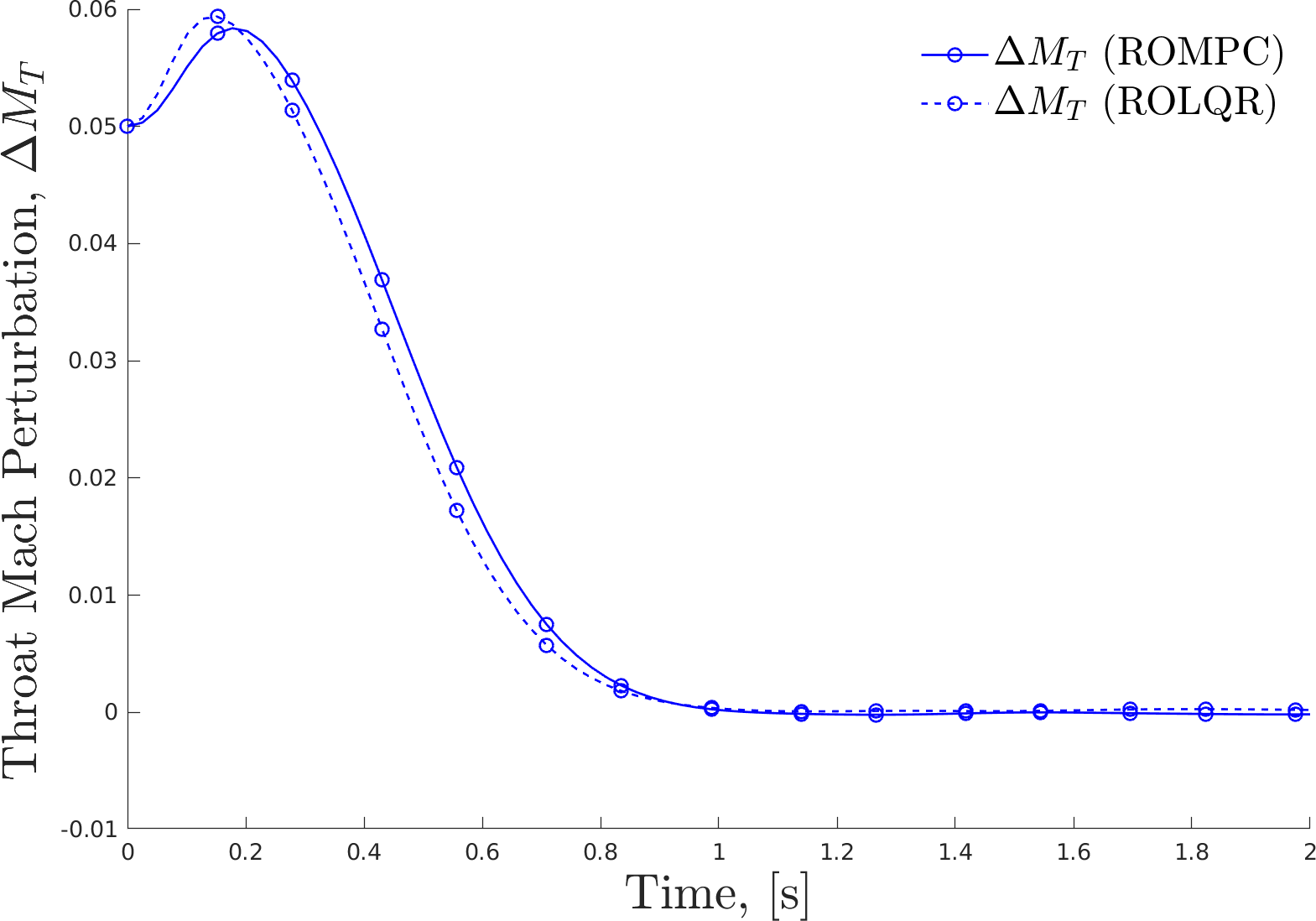}}
\caption{Simulation results for the supersonic diffuser example in Section \ref{subsec:supdiff}. This plot shows the performance output for two simulations, one with the ROMPC scheme and another with a reduced order LQR controller.}
\label{fig:z_sup_diffuser}
\end{figure}

Simulation results are shown in Figures \ref{fig:u_sup_diffuser} and \ref{fig:z_sup_diffuser}, which compare the proposed ROMPC scheme and a simple reduced order LQR controller. As can be seen in Figure \ref{fig:u_sup_diffuser}, the proposed ROMPC scheme is able to satisfy the control constraints while the LQR controller violates them. Since control constraints often represent physical actuator limits, controllers that are unable to account for such limitations may under-perform or even be unsafe.
In this simulation the initial condition is a steady state for the system where Assumption \ref{ass:inf_hor_lp} is satisfied.
To additionally highlight the motivation for the proposed approach, note that a MPC scheme designed based on the \textit{full order model} ($n^f = 11730$) is not practical due to extreme computational requirements.

\begin{figure*}[!t]
\centerline{\includegraphics[width=.9\textwidth]{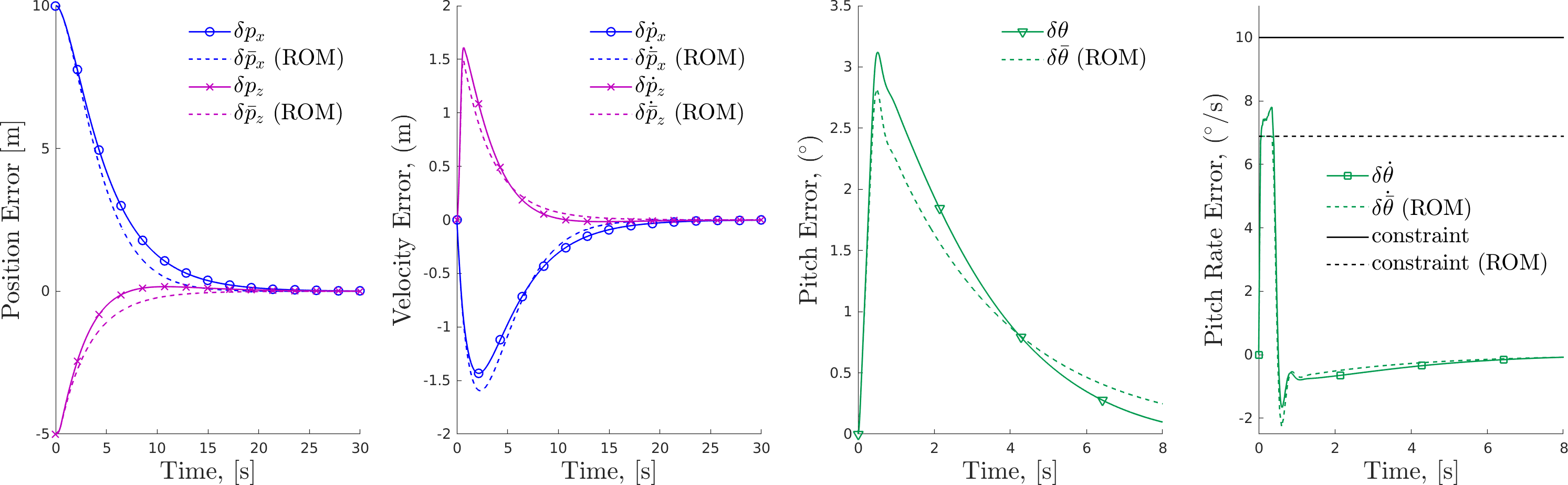}}
\caption{Simulation results for the aircraft control problem in Section \ref{subsec:aircraft}. The trajectory of the controlled aircraft is shown with solid lines and the trajectory of the simulated ROM is shown with dashed lines. Note that only the first 8 seconds of the pitch and pitch rate trajectories are shown for better detail.}
\label{fig:z_aircraft}
\end{figure*}

\subsection{Aircraft} \label{subsec:aircraft}
This example studies an automated glideslope tracking problem for a UAV in the context of autonomous aircraft carrier landing. The UAV model is expressed in continuous-time, and is a coupled fluid-structure interaction model consisting of the longitudinal rigid-body dynamics equations and a CFD model for modeling the aerodynamics, and is linearized about the equilibrium glideslope trajectory \cite{LorenzettiMcClellanEtAl2020, McClellanLorenzettiEtAl2020}. The combined rigid-body/CFD model has dimension $n^f = 998936$, has $m=2$ inputs (thrust $T$ and pitching moment $M$), and $o=p=6$ performance and measured outputs. Specifically, the outputs are the rigid-body states which represent the deviation from the glideslope trajectory: $(\delta p_x,\delta p_z)$ are positions, $(\delta \dot{p}_x, \delta \dot{p}_z)$ are velocities, and $(\delta \theta, \delta \dot{\theta})$ are the pitch angle and pitch rate. A continuous-time ROM with dimension $n=11$ is developed using POD, a popular data-driven model reduction method capable of handling extremely high-dimensional systems \cite{Antoulas2005, BennerGugercinEtAl2015}. For this particular example no external disturbances $w$ and $v$ are included to highlight the effects of the model reduction errors. Note that the extreme high-dimensionality of the coupled fluid-structure interaction model precludes the use of traditional MPC schemes.

The constraints on the performance variables and control inputs are given by $\lvert \delta p_x\rvert \leq 20 \text{m}$, $\lvert \delta p_z \rvert \leq 20\text{m}$, $\lvert \delta \dot{p}_x \rvert \leq 10$, $\lvert \delta \dot{p}_z \rvert \leq 5$, $\lvert \delta \theta \rvert \leq 10^\circ$, $\lvert \delta \dot{\theta} \rvert \leq 10^\circ/\text{s}$, $\lvert \delta T \rvert \leq 100 \text{N}$, and $\lvert \delta M \rvert \leq 50 \text{Nm}$. The controller gains $K$ and $L$ are computed using Algorithm \ref{alg:controllersynth}, and the error bounds are computed using the continuous-time method discussed in Section \ref{sec:continuoustime}. Since the system is so high-dimensional it is not practical to compute the $\Delta^{(1)}$ bound and so only the $\Delta^{(2)}$ bound is computed. The horizon $\tau = 75$ seconds is used, and the matrices defining the cost function in \eqref{eq:inf_hor_lp_e2:ct} are computed by simulating the error system with an implicit Euler scheme and leveraging sparse matrix methods. A ROM discretized with a sampling time of $0.05$ seconds is used for the computation of the error bounds and in the OCP. The computed error bounds, expressed as a percentage of the original constraint, are given in Table \ref{tab:aircraftbounds}. 

\begin{table}[htbp]
\vspace{-5pt}
\caption{Aircraft Error Bounds}
\centering
\renewcommand{\arraystretch}{1.2}
\begin{tabular}{|c|c|c|c|c|c|c|c|}
\cline{1-8}
$\lvert \delta p_x \rvert$ &  $\lvert \delta p_z \rvert$ & $\lvert \delta \dot{p}_x \rvert$ & $\lvert \delta \dot{p}_z \rvert$ & $\lvert \delta \theta \rvert$ & $\lvert \delta \dot{\theta} \rvert$ & $\lvert \delta T \rvert$ & $\lvert \delta M \rvert$ \\ \hline
 21.9 & 19.5 & 17.8 & 37.7 & 30.4 & 31.1 & 26.4 & 17.5 \\ \hline
\multicolumn{8}{p{220pt}}{Error bounds associated with the aircraft control problem in Section \ref{subsec:aircraft}. These bounds are expressed as percentages of the original constraint, are computed via the continuous-time method described in Section \ref{sec:continuoustime}, and do not include the components $\Delta^{(1)}$.}\\
\end{tabular}
\label{tab:aircraftbounds}
\vspace{-5pt}
\end{table}

Simulation results (rigid-body states) are shown in Figure \ref{fig:z_aircraft}. Trajectories of both the FOM and simulated ROM are shown, and it can be seen that the simulated ROM rides along the tightened pitch rate constraint boundary at the beginning of the trajectory. However, by using the computed error bounds the controlled aircraft still satisfies all of the true constraints. The average computation time for solving each OCP using the IBM ILOG CPLEX solver is $19.6$ milliseconds.

\section{Conclusion}
In this work a reduced order model predictive control scheme is proposed for solving constrained optimal control problems for high-dimensional systems. This approach leverages reduced order models for computationally efficient controller design, and through the proposed synthesis and analysis techniques guarantees robust stability and constraint satisfaction for the controlled high-dimensional system. Applications of particular interest include systems whose models are derived from finite approximations to infinite-dimensional systems, of which several examples are provided.

{\em Future Work:} There are several additional considerations of both theoretical and practical significance related to this work. First, an analysis on the sub-optimality of the ROMPC scheme would be of theoretical importance. Second, additional techniques for synthesizing the feedback controller gains could be considered. Third, extensions to high-dimensional nonlinear or parametric systems would be of significant interest. Additional theoretically important topics include the study of when and how model order reduction algorithms can preserve controllability and observability properties, and the extension to dynamics models
\begin{equation*}
M^fx^f_{k+1} = A^fx^f_k + B^fu_k + B^f_w w_k,
\end{equation*}
where $M^f$ is singular.

\bibliographystyle{IEEEtran}
\bibliography{main,ASL_papers}

\appendix

\begin{IEEEproof}[Proof of Lemma \ref{lem:robustconstraint}] \label{proof:robustconstraint}
Since the optimal control problem is designed to be recursively feasible and is assumed to be feasible at time $k_0$, then the simulated reduced order system trajectory will satisfy $\bar{z}_k \in \bar{\mathcal{Z}}$ and $\bar{u}_k \in \bar{\mathcal{U}}$ for all $k \geq k_0$. Therefore, since $H_z \bar{z}_k \leq b_z - \Delta_z$ and $H_ze_{z,k} \leq \Delta_z$ for all $k \geq k_0$, then $H_z \bar{z}_k \leq b_z - H_ze_{z,k}$ and finally $H_z z_k \leq b_z$ for all $k \geq k_0$. The same analysis can also be applied for $u_k$.
\end{IEEEproof}
\vspace{\baselineskip}
\begin{IEEEproof}[Proof of Theorem \ref{thm:robuststability}] \label{proof:robuststability}
Similarly to the closed-loop error dynamics \eqref{eq:edynamics}, the closed-loop dynamics of the full order system and reduced order state estimator can be written
\begin{equation*}
\xi_{k+1} = A_\epsilon \xi_k + B_\xi r_k + G_\epsilon \omega_k, \quad B_\xi = \begin{bmatrix}
-B^fK & B^f \\ -BK & B
\end{bmatrix},
\end{equation*}
where $\xi = [(x^f)^T, \: \hat{x}^T]^T$.
Therefore, writing the solution of this dynamics recursion in the form
\begin{equation*}
\xi_k = A_\epsilon^k \xi_0 + \sum_{j=0}^{k-1} A_\epsilon^{k-1-j}(B_\xi r_j + G_\epsilon \omega_k),
\end{equation*}
we can analyze the norm of $\xi$. Since $A_\epsilon$ is Schur stable there is guaranteed to exist values $M_1 \geq 1$, $\gamma_1 \in (0,1)$ such that $\lVert A_\epsilon^k \rVert \leq M_1 \gamma_1^k$ for all $k \geq 0$. Additionally, since $\mathcal{W}\times \mathcal{V}$ is a compact set by Assumption \ref{ass:compact}, there exists a constant $C$ such that $\lVert G_\epsilon \omega_k \rVert \leq C$ for all $k$. Therefore we can write
\begin{equation*}
\lVert \xi_k \rVert \leq M_1 \gamma_1^k \lVert \xi_0 \rVert + \sum_{j=0}^{k-1} M_1 \gamma_1^{k-1-j}(\lVert B_\xi r_j \rVert + C).
\end{equation*}
Further, from the exponential stability of the simulated reduced order system there exists values $M_2 > 0$ and $\gamma_2 \in (0,1)$ such that if the initial OCP is feasible then $\lVert r_k \rVert \leq M_2 \gamma_2^k$ for all $k \geq 0$. Defining $\gamma = \max\{\gamma_1, \gamma_2\} \in (0,1)$ and noting that $\sum_{j=0}^{k-1} \gamma^{k-1-j} \leq \frac{1}{1-\gamma}$ we can simplify the previous expression to
\begin{equation*}
\begin{split}
\lVert \xi_k \rVert \leq M_1 \lVert \xi_0 \rVert \gamma^k +  M_1M_2 \lVert B_\xi \rVert k\gamma^{k-1} + \frac{M_1C}{1-\gamma} . 
\end{split}
\end{equation*}
Since $\lvert \gamma \rvert < 1$, then in the limit $\lim_{k \rightarrow \infty} \gamma^k = 0$ and $\lim_{k \rightarrow \infty} k \gamma^{k-1} = 0$ and therefore 
\begin{equation*}
\lim_{k \rightarrow \infty} \lVert \xi_k \rVert \leq \frac{M_1C}{1-\gamma}.
\end{equation*}
Therefore the closed-loop system asymptotically converges to the compact set $\Xi \coloneqq \{\xi \:|\: \lVert \xi \rVert \leq \frac{M_1C}{1-\gamma} \}$.
\end{IEEEproof}
\vspace{\baselineskip}
\begin{IEEEproof}[Proof of Corollary \ref{crl:stability}]
\label{proof:stability}
In the absence of disturbances, the constant $C$ in the proof of Theorem \ref{thm:robuststability} can be chosen to be $C=0$. Therefore 
\begin{equation*}
\lim_{k \rightarrow \infty} \lVert \xi_k \rVert = 0,
\end{equation*}
and the system asymptotically converges to the origin.
\end{IEEEproof}

\vspace{\baselineskip}
\begin{IEEEproof}[Proof of Theorem \ref{thm:robusttracking}] \label{proof:robusttracking}
Consider the combined errors $\delta \xi= [(\delta x^f)^T, \delta \hat{x}^T]^T$ where $\delta x^f = x^f - x^f_\infty$ and $\delta \hat{x} = \hat{x} - \hat{x}_\infty$. The dynamics of these errors are given by 
\begin{equation*}
\delta \xi_{k+1} = A_\epsilon \delta \xi_k + B_\xi \delta r_k + G_\epsilon \omega_k,
\end{equation*}
where $\delta r= [(\delta \bar{x})^T, \delta \bar{u}^T]^T$ where $\delta \bar{x} = \bar{x} - \bar{x}_\infty$ and $\delta \bar{u} = \bar{u} - \bar{u}_\infty$. As shown in the proof of Theorem \ref{thm:robuststability}, since the simulated ROM converges exponentially to $(\bar{x}_\infty, \bar{u}_\infty)$, in the limit
\begin{equation*}
\lim_{k \rightarrow \infty} \lVert \delta \xi_k \rVert \leq B,
\end{equation*}
where $B \in \mathbb{R}$ is finite. Further, since $S^f$ and $S$ are square and full rank the combined steady state of the controller system defined by the set of points $(x^f_\infty, u_\infty)$, $\hat{x}_\infty$, and $(\bar{x}_\infty, \bar{u}_\infty)$ is unique and thus $x^f$ will converge to a vicinity of the state $x^f_\infty$. With the tracking error $\delta z^r \coloneqq z^r - r = TH^f \delta x^f$ it must also be true that
\begin{equation*}
\lim_{k \rightarrow \infty} \lVert \delta z^r_k \rVert \leq \lVert TH^f \rVert B.
\end{equation*}
Therefore the tracking variables also converge to a compact set containing the setpoint $r$.
\end{IEEEproof}

\vspace{\baselineskip}
\begin{IEEEproof}[Proof of Corollary \ref{crl:tracking}]
\label{proof:tracking}
Following the same approach as in the proof of Corollary \ref{crl:stability}, in the absence of disturbances the constant $B$ in the proof of Theorem \ref{thm:robusttracking} can be chosen to be $B=0$. Therefore $\lim_{k \rightarrow \infty} \lVert \delta z^r_k \rVert = 0$.
\end{IEEEproof}

\vspace{\baselineskip}
\begin{IEEEproof}[Proof of Proposition \ref{prop:Xbar}]
\label{proof:Xbar}
We begin by showing $\bar{\mathcal{X}}$ is compact. From Assumption \ref{ass:ctrlobsv}, the matrix $\mathcal{O} \coloneqq \begin{bmatrix} H^T & (HA)^T & \dots & (HA^{n-1})^T \end{bmatrix}^T$ is full rank. Since $\bar{x}_i = A^i \bar{x}_0 + \delta_i$ where $\delta_i = \sum_{j=0}^{i-1}A^{i-1-j}B\bar{u}_j$, the constraints $H\bar{x}_i \in \mathcal{Z}$ in \eqref{eq:Xbar} can be written as $HA^i\bar{x}_0 + H\delta_i \in \mathcal{Z}$. 
Further, the constraints of \eqref{eq:Xbar} enforce $\bar{u} \in \mathcal{U}$ and $\mathcal{U}$ is compact (Assumption \ref{ass:compact}) such that the terms $\delta_i$ are bounded. Therefore, with $\bar{i} \geq n-1$ and $\mathcal{Z}$ compact (Assumption \ref{ass:compact}), the vector $\mathcal{O}\bar{x}_0$ is bounded, and since $\mathcal{O}$ is full rank $\bar{x}_0$ is bounded as well. Thus each element $b_{\bar{x},l}$ is bounded and by choice of $H_{\bar{x}}$ the set $\bar{\mathcal{X}}$ is compact. To prove that $\bar{x}_k \in \bar{\mathcal{X}}$ for all $k \geq \underline{k}$ we simply note that by the proposition assumptions the constraints of \eqref{eq:Xbar} are satisfied for all $k \geq \underline{k}$ and therefore $b_{\bar{x}}$ is a valid upper bound on $H_{\bar{x}}\bar{x}$.
\end{IEEEproof}

\vspace{\baselineskip}
\begin{IEEEproof}[Proof of Proposition \ref{prop:CrandCw}] \label{proof:CrandCw}
By Proposition \ref{prop:Xbar} and Assumption \ref{ass:compact} the sets $\bar{\mathcal{X}}$, $\mathcal{U}$, $\mathcal{V}$, and $\mathcal{W}$ are compact, which guarantees $C_r$ and $C_\omega$ are finite. Additionally, by the bounded disturbance assumption it is guaranteed that $\omega_k \in \mathcal{W}\times \mathcal{V}$ for all $k$ and therefore $\lVert G_\epsilon \omega_k \rVert_G \leq C_\omega$ for all $k$. Similarly, by the proposition assumptions $r_k \in \bar{\mathcal{X}}\times \mathcal{U}$ for all $k\geq \underline{k}$ such that $\lVert B_\epsilon r_k \rVert_G \leq C_r$ for all $k \geq \underline{k}$.
\end{IEEEproof}

\vspace{\baselineskip}
\begin{IEEEproof}[Proof of Theorem \ref{thm:robustconstraint}] \label{proof:robustconstraintthm}
By design, under Assumptions \ref{ass:compact}-\ref{ass:inf_hor_lp} and from Propositions \ref{prop:Xbar} and \ref{prop:CrandCw} the bounds $\lVert \epsilon_k^{(1)} \rVert_G \leq \Delta^{(1)}$ and $\theta^T\epsilon_k^{(2)} \leq \Delta^{(2)}(\theta)$ hold for any $\theta$ and for all $k \geq k_0$. Therefore by definition $H_z \delta_{z,k} \leq \Delta_z$ and $H_u \delta_{k,u} \leq \Delta_u$ for all $k \geq k_0$. The final result therefore follows from Lemma \ref{lem:robustconstraint}.
\end{IEEEproof}

\begin{IEEEbiographynophoto}{Joseph Lorenzetti} is a Ph.D. candidate working in the Autonomous Systems Laboratory at Stanford University. He earned a M.S. in Aeronautics and Astronautics from Stanford University in 2018, and a B.S. in Aeronautical and Astronautical Engineering from Purdue University in 2015. His main research focus is on computationally efficient and high-performing algorithms for controlling infinite-dimensional systems, with applications to autonomous aircraft control and soft robotics. Joseph is currently supported by the Department of Defense (DoD) through the National Defense Science and Engineering Fellowship (NDSEG) Program.
\end{IEEEbiographynophoto}

\begin{IEEEbiographynophoto}{Andrew McClellan} graduated with a B.S. in Physics from Brigham Young University in 2016 and a M.S. in Aeronautics and Astronautics from Stanford University in 2018. He currently is a Ph.D. candidate at Stanford in the Department of Aeronautics and Astronautics. 
His research focuses on producing projection-based reduced order models for use in model predictive control applications. This primarily involves determining the training procedure used to produce solution snapshots, such that the resulting snapshots accurately cover the solution domain of interest and produce an efficient compression. Andrew is supported by the Stanford Graduate Fellowship.
\end{IEEEbiographynophoto}

\begin{IEEEbiographynophoto}{Charbel Farhat} is the Vivian Church Hoff Professor of Aircraft Structures, Chairman of the Department of Aeronautics and Astronautics, and Director of the KACST Center of Excellence for Aeronautics and Astronautics at Stanford University. His research interests are in computational engineering sciences for the design and analysis of complex systems in aerospace, mechanical, and naval engineering. He is a Member of the National Academy of Engineering, a Member of the Royal Academy of Engineering (UK), a designated ISI Highly Cited Author, and a Fellow of AIAA, ASME, IACM, SIAM, USACM, and WIF. He is also the recipient of many professional and academic distinctions including the Gordon Bell Prize and the Sidney Fernbach Award from IEEE.
\end{IEEEbiographynophoto}

\begin{IEEEbiographynophoto}{Marco Pavone} is an Associate Professor of Aeronautics and Astronautics at Stanford University, where he is the Director of the Autonomous Systems Laboratory and Co-Director of the Center for Automotive Research at Stanford. Before joining Stanford, he was a Research Technologist within the Robotics Section at the NASA Jet Propulsion Laboratory. He received a Ph.D. degree in Aeronautics and Astronautics from the Massachusetts Institute of Technology in 2010. His main research interests are in the development of methodologies for the analysis, design, and control of autonomous systems, with an emphasis on self-driving cars, autonomous aerospace vehicles, and future mobility systems. He is a recipient of a number of awards, including a Presidential Early Career Award for Scientists and Engineers (PECASE), an ONR YIP Award, an NSF CAREER Award, and a NASA Early Career Faculty Award. He was identified by the American Society for Engineering Education (ASEE) as one of America’s 20 most highly promising investigators under the age of 40. He is currently serving as an Associate Editor for the IEEE Control Systems Magazine.
\end{IEEEbiographynophoto}

\end{document}